%% file: 2M_ZOA_MN.tex
\documentclass[usenatbib]{mn2e}

 \usepackage{times}

\usepackage{cite}
\usepackage{epsfig}
\usepackage{color}
\usepackage{lscape}
\usepackage{longtable}
\usepackage{journal_names}
\usepackage{natbib}
\usepackage{float}

\newcommand{\HI}{\mbox{H\,{\sc i}}}

\newcommand{\HIit}{\mbox{H\hspace{0.155 em}{\footnotesize \it I}}}

\newcommand{\K}{$K_s$}

\newcommand{\am}[2]{$#1'\,\hspace{-1.7mm}.\hspace{.0mm}#2$}

\newcommand{\masq}{\mbox{mag~arcsec$^{-2}$}}
\newcommand{\kms}{\mbox{km\,s$^{-1}$}}
\newcommand{\nan}{Nan\c{c}ay}

\newcommand{\kmsMpc}{\mbox{ km~s$^{-1}$~Mpc$^{-1}$}}
\newcommand{\KMS}{\mbox{$\frac{\rm {km}}{\rm {s}}$}}
\newcommand{\JYKMS}{\mbox{$\frac{\rm {Jy km}}{\rm {s}}$}}
\newcommand{\Msun}{M$_\odot$}
\newcommand{\Lsun}{L$_\odot$}

\newcommand{\LsunK}{\mbox{L$_{\odot, \rm K}$}}

\newcommand{\FHI}{\mbox{$F_{\rm {HI}}$}}

\newcommand{\Koc}{$K^{\rm o,d}_{\rm s}$}
\newcommand{\koc}{$K^{\rm o,d}_{\rm s}$}
\newcommand{\rkc}{\mbox{$r_{K_s}^{\rm o,d}$}}
\newcommand{\hkoc}{$(H-K_{\rm s})^{\rm o,d}$}
\newcommand{\jkoc}{$(J-K_{\rm s})^{\rm o,d}$}

\newcommand{\MHI}{$M_{\rm {HI}}$}
\newcommand{\MHIstar}{$M^{*}_{{\rm {HI}}}$}
\newcommand{\lMHI}{$\log(M_{\rm {HI}})$}

\newcommand{\Mdyn}{$M_{\rm {dyn}}$}
\newcommand{\lMdyn}{$\log(M_{\rm {dyn}})$}
\newcommand{\Mbar}{$M_{\rm {bar}}$}
\newcommand{\Mbardyn}{$M_{\rm {bar}}/M_{\rm {dyn}}$}
\newcommand{\lMbardyn}{$\log({M_{\rm {bar}}\over M_{\rm {dyn}}})$}

\newcommand{\MHILK}{\mbox{$M_{\rm {HI}}/L_{\rm K}$}}
\newcommand{\MHILKc}{\mbox{$M_{\rm {HI}}/L_{K_c}$}}

\newcommand{\LKc}{\mbox{$L_{\rm K_c}$}}
\newcommand{\lLKc}{\mbox{$\log(L_{\rm K_c}$})}

\newcommand{\VHI}{\mbox{${V}_{\rm {HI}}$}\,}
\newcommand{\Vopt}{\mbox{${V}_{\rm {opt}}$}\,}
\newcommand{\vopt}{\mbox{${V}_{\rm {opt}}$}\,}
\newcommand{\vrot}{\mbox{$V_{\rm {rot}}$}\,}
\newcommand{\Wfifty}{\mbox{${W}_{\mathrm50}$}}
\newcommand{\Wtwenty}{\mbox{${W}_{\mathrm20}$}}
\newcommand{\wfi}{\mbox{$W_{\mathrm 50}$}}
\newcommand{\wtw}{\mbox{$W_{\mathrm 20}$}}
\newcommand{\Vfifty}{\mbox{$V_{\mathrm 50}$}}
\newcommand{\vfi}{\mbox{$V_{\mathrm 50}$}}

\title[{The \nan\ \HIit\, ZoA survey of 2MASS bright galaxies}]{The \nan\ H{\Large \bf I} Zone of Avoidance survey of 2MASS bright galaxies}

\author[Ren\'ee C. Kraan-Korteweg et al.]
{Ren\'ee C. Kraan-Korteweg$^{1}$\thanks{E-mail: kraan@ast.uct.ac.za}, 
Wim van Driel$^{2,3}$, 
Anja C. Schr\"oder$^{4}$,
Mpati Ramatsoku$^{1,5}$  
\newauthor 
and
Patricia A. Henning$^{6}$ \\ 
$^{1}$Department of Astronomy, University of Cape Town, Private Bag X3, Rondebosch 7701, South Africa\\
$^{2}$GEPI, Observatoire de Paris, PSL Research University, CNRS, Universit\'e Paris Diderot, 5 place Jules Janssen, 92190 Meudon, France\\
$^{3}$Station de Radioastronomie de \nan, Observatoire de Paris, CNRS/INSU USR 704, Universit\'e d'Orl\'eans OSUC, 
route de Souesmes, 18330 \nan, France \\
$^{4}$South African Astronomical Observatory, PO Box 9, Observatory 7935, Cape Town, South Africa\\
$^{5}$INAF- Osservatorio Astronomico di Cagliari, Via della Scienza 5, I-09047 Selargius (CA), Italy\\
$^{6}$Department of Physics and Astronomy, University of New Mexico, Albuquerque, NM 87131, USA
}

\begin{document}

\date{Accepted....... ;}

\pagerange{\pageref{firstpage}--\pageref{lastpage}} \pubyear{2018}

\maketitle

\label{firstpage}

\begin{abstract}
To complement the 2MASS Redshift Survey (2MRS) and the 2MASS Tully-Fisher survey (2MTF) a search for 21cm \HI\ line emission of 2MASS bright galaxy candidates has been pursued along the dust-obscured plane of the Milky Way  with the 100m \nan\ Radio Telescope. For our sample selection we adopted an isophotal extinction-corrected \K -band magnitude limit of $K_{\rm s}^{\rm o} = 11\fm25$, corresponding to the first 2MRS data release and 2MTF, for which the 2MASX completeness level remains fairly constant deep into the Zone of Avoidance (ZoA). About one thousand galaxies without prior redshift measurement accessible from \nan\ (Dec $> -40\degr$) were observed to an rms noise level of $\sim 3$\,mJy for the velocity range $-250$ to $10\,600$\,\kms. This resulted in 220 clear and 12 marginal detections of the target sample. Only few detections have redshifts above 8000\,\kms\ due to recurring radio frequency interference (RFI). 
A further 29 detections and 6 marginals have their origin in non-target galaxies in the telescope beam. The newly detected galaxies are on average considerably more \HI-rich (mostly $10^9 - 10^{10}$M$_\odot$) compared to systematic (blind) \HI\ surveys. The \HI\ detections reveal various new filaments crossing the mostly uncharted northern ZoA (e.g. at $\ell \sim 90\degr, 130\degr, 160\degr$), whilst consolidating galaxy agglomerations in Monoceros and Puppis ($\ell \sim 220\degr, 240\degr$). Considerably new insight has been gained about the extent of the Perseus-Pisces Supercluster through the confirmation of a ridge ($\ell \sim 160\degr$) encompassing the 3C129 cluster that links Perseus-Pisces to Lynx, and the continuation of the second Perseus-Pisces arm ($\ell \sim 90\degr$) across the ZoA.

\end{abstract}


\section[]{Introduction} \label{intro} 

Our understanding of the large-scale structures and associated dynamics in  the nearby Universe is increasingly being refined thanks to the advances of  larger and more systematic redshift surveys \citep{Jarrett2004,jones2009,huchra12,tempel2014} and subsequent peculiar velocity surveys \citep{springob2014,scrimgeour2016,springob2016}.  Despite all these efforts, the overall bulk flow resulting from these studies can still not be fully reconciled with the dipole observed in the Cosmic Microwave Background \citep{Fixsen1996}, nor is there consensus about the volume which gives rise to the bulk flow \citep[e.g.][]{hudson2004, erdogdu2006, kocevski2006, carrick2015, hoffman2015}. This has left a gap in our knowledge of the local large-scale structure and therefore considerable uncertainties in our understanding of the dynamics in the local Universe, cosmic flow fields and the dipole. 

One of the major limitations remains the lack of data on galaxies in the so-called Zone of Avoidance (ZoA), caused by dust extinction and star crowding at low Galactic latitudes, see e.g. \citealt{KKLahav2000, KK2005} for reviews. This zone is even wider in Galactic latitude for ``3-dimensional'' (redshift) surveys, because optical redshifts are increasingly difficult to obtain for dust-obscured galaxies towards the Galactic plane. 

The ZoA is known to obscure and bisect major parts of dynamically important structures such as the Great Attractor \citep{Dressler1987, KK1996, radburn2006, woudt2008, hizoa2016}, the Perseus-Pisces Supercluster \citep{giovanelli1982, focardi86,hauschildt87} and the Local Void \citep{Tully2008, KKLV2008}. Moreover, the ZoA could still surprise us with completely unknown structures as postulated, for instance, by \citet{loeb08} -- and which indeed are still being found, as demonstrated by the recently discovered Vela Supercluster \citep{rkk_salt_15,KK2017}. 

A vast improvement was achieved with the near-infrared 2-Micron All Sky Survey  \citep[2MASS;][]{skrutskie06} from which an extended source catalogue of 1.6 million objects was extracted, \citep[2MASX;][]{jarrett03}, which is uniform over the sky and much less affected by dust extinction than optical whole-sky surveys. It is complete to the magnitude limit of $K_s < 13\fm5$, and consists mostly are galaxies, apart from an increasing number of (extended) Galactic sources close to the Galactic Plane. Around the Galactic Bulge, 2MASX suffers some incompleteness though due to the high  stellar density. According to \citet{Jarrett+2000a} galaxy counts start to drop where stellar density levels reach log~$N({\rm SD}_{K_s < 14\fm0}) = 3.6$ per sq degree, hence  from about $-90\degr \la \ell \la +90\degr$  for latitudes below $|b| \la 5\degr$. The Milky Way becomes completely opaque to galaxies where this level reaches log~$N=4.5$, leading to a broadening of the near-infrared 2MASX-ZoA to $|b| \la 8\degr$ for about $\Delta \ell \pm30 \degr$ around the Galactic Centre (GC). However, the Galactic Anticentre region ($90\degr < \ell < 270\degr$) -- hence all of the northern ZoA -- is not affected, at least not for the brightest 2MASX galaxies.
 
It is therefore not surprising that 2MASX was used as a basis for a systematic redshift follow-up, launched over a decade ago by John Huchra, to arrive at a homogeneous 'whole-sky' 2MASS Redshift Survey (2MRS) for large-scale structure studies of the nearby Universe. Its release came in batches, starting in 2005 with the first on-line release \citep[see e.g.][]{huchra05}, a survey of $\sim$\,23\,000 galaxies which is complete to an extinction-corrected magnitude of $K^o \le 11\fm25$. The second data release goes half a magnitude deeper ($K^o \le 11\fm75$) and contains $\sim$~44\,500 galaxies \citep{huchra12}. Both versions however did not target 2MASX galaxies in the innermost ZoA ($|b| < 5\degr$; extending to $|b| < 8\degr$ around the Galactic Bulge) because of the inherent difficulties in getting good signal-to-noise (SNR) optical spectra for these heavily obscured galaxies. 

This gap in the large-scale structure maps of the nearby Universe has partly improved through systematic blind \HI\ surveys, which do not suffer foreground extinction, e.g. HIPASS at the 64m Parkes radio telescope for the southern sky \citep{Meyer2004,2004BGC}, its northern extension to Dec$=+25\degr$ \citep{Wong2006}, and the on-going \HI\ 100m Effelsberg radio telescope \HI-Survey (EBHIS) for the remaining northern sky \citep{kerp11}. But these surveys are shallow (rms $\sim17-25$ mJy), and sparsely sampled in comparison to optical and NIR whole-sky galaxy surveys in terms of detections per unit sky area ($\sim$ 0.1 gal/sq degree) and the volume to which they are sensitive.

To improve on this situation a deeper \HI\ survey (rms $\sim 6 $mJy), focused specifically on the most opaque part of the ZoA ($|b|\le 5\degr$), was performed with the Parkes radio telescope: HIZoA encompasses the southern ZoA \citep{hizoa2016} and its northern extension ($\Delta \ell = 16\degr$) \citep{Donley2005}. But no similar systematic \HI-survey exists for the remainder of the northern ZoA, apart from selected regions pursued with the 305m Arecibo telescope. These are the shallow ALFA ZoA survey \citep{henning10} of the inner Galaxy, at $|b| < 10\degr$ (rms $\sim 5$\,mJy), and the deep ALFA ZoA surveys \citep{henning08} for both the inner and outer Galaxy ($|b| < 5\degr$, with an rms $\sim 1$mJy). 

This leads to a significant part of the northern ZoA that has not been surveyed in velocity space in any systematic way, in particular the ZoA range in between the declination strips accessible from Arecibo ($70\degr \la \ell \la 180\degr$). The latter is particularly relevant because it is home to the Perseus-Pisces Supercluster (PPS), a dominant structure of the nearby Universe, which crosses the ZoA at two locations. The PPS has therefore never been fully charted. Another structure, the Supergalactic Plane crosses the GP in this part unexplored part of the ZoA as well, while other unknown structures may have gone completely unnoticed.

Following the success of our pilot study using the \nan\ (NRT) and the Arecibo radio telescopes to survey 197 2MASS galaxy candidates  in \HI\ \citep{vandriel09}, we decided in 2009 to fill in the 2MRS ZoA gap -- in collaboration with the late John Huchra -- and pursue a targeted \HI\ survey of all the bright ZoA 2MASX galaxies that had no prior redshift information, and can be reached with the NRT, i.e. have declinations $\delta > -40\degr$. This implies a ZoA longitude range of $\Delta \ell = 280\degr$ covering $-20\degr \la \ell \la 260\degr$. To be compatible with 2MRS, our completeness limit was set $K^o_s < 11\fm25$ {\sl after} corrections for foreground extinction (described in detail in Sect.~\ref{sample}). This limit corresponds to an optical $B$-band limit of $13\fm25$ for a typical spiral colour of $(B-K) \simeq 2\fm0$ \citep[e.g.][]{jarrett03}, and is therewith comparable to the completeness limit of most optically selected nearby whole-sky surveys.

The resulting \nan\ \HI\ ZoA target list contains about one thousand 2MASX galaxies  within the Galactic latitude range $|b| < 10\degr$, that had no prior redshift and were accessible from \nan. Most are located within $|b| < 5\degr$. This paper focuses on the NRT \HI-observations pursued between 2009\,--\,2015, the resulting detection rates, galaxy properties, and a description of the new large-scale structures uncovered by the \nan\ \HI-ZoA survey.

It should be noted that the results presented here form part of a much larger effort to (a) map the large-scale structures across the full $360\degr$-circle of the ZoA, and (b) to subsequently determine the peculiar velocity fields and underlying mass-density field for an adequately selected subsample of inclined spiral galaxies through a near-infrared (NIR) Tully-Fisher (TF) analysis optimized for the ZoA (Said et al. 2015). The latter will complement the 2MASS Tully-Fisher Survey (2MTF; e.g. {\citealt{Masters+2008, Hong2013, Hong2014, springob2016, Howlett2017}}) which excludes the inner ZoA ($|b| > 5\degr$), since our data is based on a newly defined catalogue of 2MASS bright galaxies complete to the same extinction-corrected magnitude limit of $K^o_s \le 11\fm25$ for the whole ZOA within $|b| < 10\degr$ \citep[2MZOAG;][Paper I]{Schroeder2018}.

To achieve this, various other programs have been launched, that will be presented separately. These include an extension of the \nan\ ZoA 2MASX target list to the more southern galaxies ($\delta < -40\degr$) using the Parkes radio telescope. \HI\ observations of potential TF galaxies that have optical redshifts only, as well as of re-observations of \HI-detections whose signal-to-noise ratio (SNR) were too low to allow robust line-width parameterization \citep{Said16HI}. The final step will be an application of the NIR TF relation optimized for the ZoA \citep[see][]{Said15TF}.  

An independent approach towards a southern ZoA peculiar velocity analysis  (Said et al., in prep.) consisted in a dedicated follow-up NIR imaging survey \citep{Said16NIR} of all the \HI\ detections from the systematic (blind) \HI\ survey HIZoA \citep{hizoa2016}, hence the other way around compared to starting with a NIR selected sample.
The two methods are complementary to each other in the sense that HIZoA has excellent coverage over the Galactic Bulge area where 2MASS galaxies are lacking, whereas the 2MASS bright galaxy sample is very complete over the remainder of the ZoA, ensuring coverage along the whole Galactic equator.

This paper is dedicated to reducing the 2MRS 'redshift ZoA' using the \nan\ radio telescopes and studying the newly unveiled large-scale structures, in particular in the mostly unexplored region between $70\degr \la \ell \la 180\degr$. Its structure is as follows: the selection of the 2MASS ZoA galaxy candidates observed in \HI\ is described in Sect.~2, the observations and data reduction procedures in Sect.~3, whilst the properties of the \HI-detected and non-detected galaxies are detailed in Sect.~4. Section~5 follows with a description of the global properties of the NRT detections -- a Hubble constant of $H_{\mathrm 0} = 75~$\kmsMpc was used for their derivation. The newly discovered large-scale structures are discussed in detail in Sect.~6, ending with a summary in Sect.~7.

\section[]{Sample selection} \label{sample} 
Our first 2MASX ZoA target list was provided in 2009 by John Huchra. It contained 2MASX galaxies within the Galactic latitude range $|b| < 10\degr$ that then had no prior redshift and were accessible with the NRT ($\delta > -40\degr$). It hence went wider than the 2MRS exclusion region: $|b|< 8\degr$ around the Galactic Bulge ($\ell < \pm 30\degr$) and $|b| < 5\degr$ elsewhere. The higher latitude range was motivated by the much lower fraction of completeness of the 2012 2MRS catalogue \citep{huchra12} at the lowest latitudes ($\sim$ 17.8\% for 2MRS galaxies below $|b| < 10\degr$) compared to the overall superb completeness in redshift of the full catalog of 97.6\%. And although the incompleteness fraction is significantly lower for the magnitude limit of the first data release ($K^o_s \le 11\fm25$), that was not the case when we started out with this project in 2009. Moreover, the expectations were that radio observations at the 21\,cm line might well have better success in capturing the redshifts of many of these optically obscured low-latitude 2MRS galaxies -- at least for the ones containing gas.

The resulting \nan\ observing list contained close to a thousand galaxies (for $K^o_s \le 11\fm25$, $\delta > -40\degr$), excluding the roughly 350 2MASX ZoA galaxies in Huchra's private redshift catalogue, the so-called zcat, that he had compiled for latitudes below $|b| < 5\degr$.

During the course of our observing program it became clear that the original ZoA data set needed to be adjusted, particularly -- but not exclusively -- for the $|b| < 5\degr$ band. The reason was that we did encounter low latitude 2MRS galaxies that were not in the original Huchra 2MRS target and ZoA lists. Moreover, we also found 2MASX objects that were not classified as galaxies, despite redshifts confirming their reality. As such it was decided to redefine the 2MASS galaxy ZoA list from scratch for the latitude range $|b| < 10\degr$ over the full band of the Milky Way. 

The resulting 2MASS ZoA Galaxy catalogue (2MZOAG) is described in detail in a separate paper \citep[][Paper I]{Schroeder2018} and forms the basis of all our further 2MASS ZoA studies, including a future analysis of the variation of the foreground extinction using the NIR colours of the ZoA galaxies themselves. 

For the purpose of this paper it suffices to say that the selection was based on the isophotal magnitude $K_s$-band magnitude, $K_{\rm 20}$, not the $H$-band as for the 2MRS. We furthermore worked with isophotal magnitudes rather than total magnitudes because they are more robust when  selecting galaxies that may have suffered some level of obscuration \citep{Said15TF}. A further advantage of redefining a low-latitude NIR galaxy catalogue was the availability of vastly increased multi-wavelength sky-survey imaging data for the galaxy verification process, such as UKIDSS, VISTA, and WISE, in addition to the 2MASS $J, H, K$ images, and the previously existing $B, R, I$ images of the Digitized Sky Surveys. This made the identification process more secure. We furthermore took account of the foreground extinction $E(B-V)$ at the actual position of the source \citep{schlegel98}, to ensure that  all sources up to the {\sl extinction-corrected isophotal} $K_{\rm 20}$-band magnitude limit of $K^o_s \le 11\fm25$ entered the 2MZOAG. This did take into account the {\sl additional} dimming due to the loss of the other low surface-brightness areas of a galaxy \citep{Riad2010b} over and above the extinction $A_{\rm K}$. This is relevant when trying to arrive at a complete NIR whole-sky survey that includes the ZoA. We added a few further galaxies to our target list that lie at higher latitudes. These are also located in high extinction regions, and  were excluded in 2MRS and 2MTF because of their additional sample criterion of $E(B-V) < 1\fm0$. 

\begin{figure*}
\begin{center}
\includegraphics[width=.98\textwidth,angle=0]{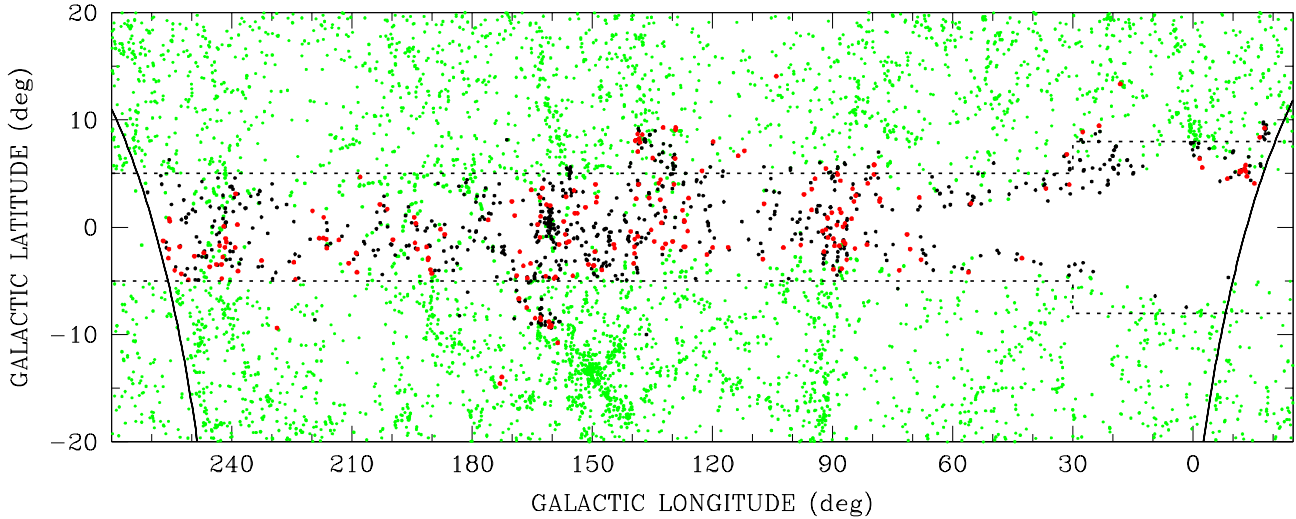}
	\caption{Distribution in Galactic coordinates of the 2MASX bright ($K\degr \le 11\fm25$) galaxy sample accessible with the NRT ($\delta > -40\degr$; black curves). The dotted black curve marks the nominal latitude limit of the 2MRS. Green dots mark galaxies with existing 2MRS and 2MZoA redshifts, the red and black dots our NRT target sample of galaxies without  redshifts ($N = 1003$), of which 23\% were detected (red) with the NRT down to the survey sensitivity limit of rms = 3~mJy. The increasing 2MASS galaxy incompleteness in the inner ZoA towards the Galactic center region due to star crowding is notable from $\ell < 80\degr$. 
    }
	\label{fig:sample}
\end{center}
\end{figure*}

In conclusion, our final \nan\ target sample consists of 1003 galaxies at low Galactic latitudes. The distribution of all targets is displayed in Fig.~\ref{fig:sample}; it shows the survey area in Galactic coordinates for the longitude range accessible with the NRT (given by the black curves). The green dots mark the 2MASX galaxies that have published redshifts, while the red and black dots indicate the target sample -- red identifies the \HI\ detected galaxies and black the non-detections (see Sect.~\ref{results} for details). The displayed 2MASS galaxies are based on a combination of the 2MZOAG catalogue for $|b| < 10\degr$, and the 2MRS galaxies for $|b| \ge 10\degr$, both for $K^o_s \le 11\fm25$. 

\begin{figure}
\includegraphics[width=.48\textwidth,angle=0]{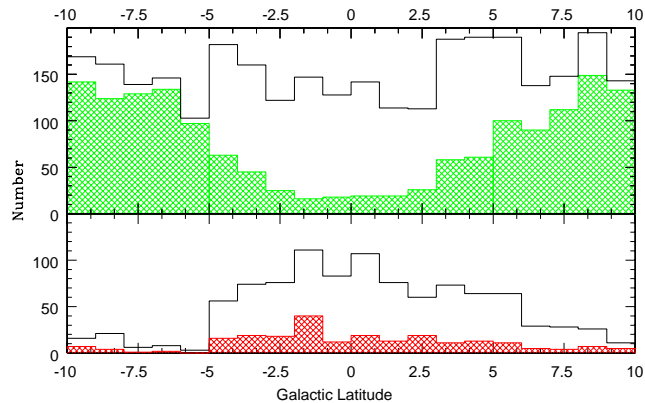}
	\caption{Histogram of the 2MASX ZoA bright galaxies with $|b| < 10\degr$ for the NRT visibility range $\delta > -40\degr$. Top panel: the black histograms demarcate all 2MZOAG galaxies, green are the ones with redshifts in 2MRS \citep{huchra12} complemented by the low latitude zcat compilation. Bottom panel: the \nan\ target sample (all observed), where the red histogram highlights the \HI\ detections (see Sect.~\ref{results}).}
    \label{fig:histsample}
\end{figure}

To demonstrate the success in bridging the 2MASX bright galaxy distribution across the ZoA we present Fig.~\ref {fig:histsample}, which displays in the top panel a histogram of the distribution of the 2MZOAG catalogue as a function of Galactic latitude ($|b| < 10\degr$) for the \nan\ survey declination range (see Paper I, for a description of the full 2MZOAG). Apart from fluctuations due to large-scale structure, hardly any dip is notable in the galaxy counts at the lowest latitude, despite part of the Galactic Bulge area being included in the survey sample (see Fig.~\ref{fig:sample}). The green histogram presents the 2MRS redshifts for $|b| > 5\degr$ -- as published in \citet{huchra12}  -- with the addition of the low-latitude zcat-compilation for $|b| < 5\degr$. 

The difference between the black and green histograms formed the basis of our NRT target selection. The final and fully observed survey sample is given by the black histogram in the lower panel (as for Fig.~\ref{fig:sample}, the \HI-detections are identified by the red histogram). Note, however, that the sum of the green histogram and the target sample do not entirely add up. During the course of the observations more redshifts became available, while a detailed literature search also revealed further published redshifts \citep[2MZOAG;][]{Schroeder2018}.

\section[]{Observations and data reduction} \label{obs} 
This section describes the actual NRT observing and data reduction procedures of the 1003 galaxy sample, followed in Sect.~\ref{results} by a presentation and analysis of the \HI- detections and non-detections. The NRT is a 100 meter-class radio telescope, with its 6900\,m$^2$ collecting area. It is a meridian transit-type instrument, see e.g. \citet{vandriel16} for further details on the instrument and data reduction. Due to the E-W elongated shape of the mirrors, some of its characteristics depend on the observed declination. Its HPBW is \am{3}{5} in right ascension, independent of declination,  while in declination it is 23$'$ for $\delta<20^{\circ}$, rising to 32$'$ at  $\delta = 71^{\circ}$ \citep[see also][]{matthews00}. The instrument's sensitivity follows the same geometric effect and decreases correspondingly with declination. The typical minimum system temperature is 35~K. 

Flux calibration is determined through regular measurements of a cold load calibrator and periodic monitoring of strong continuum sources by the \nan\ staff. For verification and monitoring of the standard, regular NRT flux scale calibration of continuum sources, we also regularly observed \HI\ line flux calibrator galaxies measured at Arecibo by \citet{oneil04HI}. From the 51 NRT measurements we derived a ratio between our total line fluxes and the literature values of 0.91$\pm$0.08  (see \citealt{vandriel16} for details on the flux scale comparison). Note that we did not use these calibrator data to rescale the \nan\ line fluxes presented in the next section.

We used an auto-correlator set-up of 4096 channels in a 50~MHz bandpass, with a channel spacing of 2.6~\kms\ and a velocity coverage of $-250$ to 10\,600~\kms\ (or 1370 to 1421\,MHz in frequency). The data were taken in position-switching mode, with an on-source integration time of 40~seconds per ON-OFF pair. The observations were made in the period of July 2009 -- December 2015, using a total of about 1800 hours of telescope time.  
Our aim was to keep integrating a source until an rms noise level of 3~mJy was reached, at a velocity resolution of 18~\kms, unless a sufficiently high signal-to-noise ratio profile was detected before the target rms value was reached.

We used standard NRT software to average the two receiver polarizations, perform the declination-dependent conversion from system temperature to flux density, fit polynomial baselines, smooth the data to a velocity resolution of 18~\kms, and ultimately convert radial velocities to the optical, heliocentric system.

We also used standard NRT software to mitigate relatively strong Radio Frequency Interference (RFI), see \citet{monnier03c} for technical details. This allows us to identify unwanted time-variable RFI, and prevents mistaking it for genuine cosmic \HI\ line signals. Removing such RFI but preserving underlying galaxian \HI\ line signals is beyond the scope of the hard- and software used. The main sources of RFI are persistent, narrow terrestrial radars at velocities between $\sim$ 8000--10,000~\kms, and a much broader, but intermittent, GPS signal centred on $\sim$8300\,\kms. The number of radars varies between observations but they are rarely all absent and in general so densely packed in velocity as to make the detection of galaxian \HI\ lines in between them impossible. In the rare cases where the intermittent GPS signal is so strong that it causes baseline ''ringing'' over a broad velocity range, the affected ON/OFF cycles are not used for further data analysis; otherwise the GPS signal is flagged but not removed.

Spectra of 34 sources appeared affected by a well-known instrumental baseline ripple \citep[e.g.][]{wilson09} which can significantly increase the rms noise level. The ripple, related to the presence of a strong continuum source, is caused by reflected radiation in the telescope which forms a standing wave with a wavelength of about 115 \kms. It is characterized by a narrow peak always at the same position in an FFT deconvolution of a spectrum, which can be effectively identified and removed; a second FFT transform then results in a de-rippled line spectrum. This ``derippling'' was performed by M.D. Lehnert, using a Python routine written by him (see \citealt{butcher16})

In our adjudication of detections, we divided them into four categories: clear, marginal, possible and non-detections. This is based on independent visual inspections of the spectra as well as on the S/N parameter \citep[cf.] []{saintonge07}.

\section[]{Resulting Data} \label{results} 
With a total of 1013 pointings we observed 1003 2MASX galaxies above our extinction-corrected magnitude limit. This number includes 12 galaxies that lie in pockets of higher extinction of $E(B-V) \geq 1\fm0$ but lie above the $|b| > 10\degr$ survey border, and  were therefore excluded from the original 2MRS (or TMF) samples. They are marked with the flag (e) in the Tables discussed below. 

Of the observed sample galaxies, 220 are clear and 10 marginal detections (4 clear and 1~marginal lie above the ZoA). This constitutes a detection rate of 22\%, which is satisfactory given that (i) no pre-selection on type or likelihood of gas content was attempted, and (ii) the ZoA declination strips within reach of the Arecibo telescope had been covered extensively by \citet{pantoja97}. These consisted of pointed \HI\ observations of galaxies identified in a dedicated search on sky survey plates, hence many are NIR-bright. 

The 22\% detection rate is practically identical to the one obtained in our pilot project \citep{vandriel09} for the 2MASX galaxies observed with the NRT (not counting the deeper Arecibo targets), which reached similar rms levels. 

A further 29 clear and 6 marginal detections were identified in spectra which -- after careful scrutiny -- turned out to be the signal of another galaxy in the NRT beam. They are henceforth referred to as non-target detections and are listed in a separate table (Table 2). They were not used in any of the subsequent analyses. 

Nine galaxies that were not in our magnitude limited target list, but were observed {\sl and detected} in the course of the program to confirm the \HI-flux of nearby sources and to avoid possible confusion issues with the actual target. Of these 6 were clearly deteced and 3 marginally. The tenth pointing of a non-sample galaxy confirmed the original sample galaxy detection. These 10 galaxies have not been used either in any of the subsequent analyses. 

The remaining 773 galaxies were not detected (Sect.~\ref{nondets}). This number includes 12 galaxies were a hint of a signal is seen. Despite increased integration times leading to rms-values below our sensitivity limit, the S/N remained too low for a credible detection. We call these "possibles", to differentiate them from the marginals whose signals, although weak, look realistic.

\begin{figure}
\begin{center}
\includegraphics[scale=0.85,angle=0]{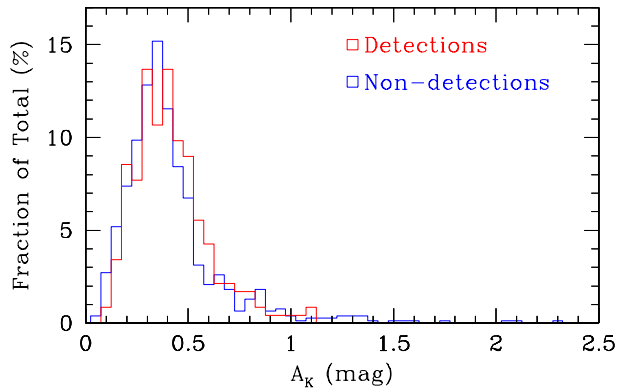}
	\caption{Distribution of extinction $A_{\rm K}$ of the in \HI\ detected galaxies (red; $N = 230$) versus the non-detections (blue; $N = 773$), both as a fraction of their respective total. As expected, the detection rate shows no dependence on foreground extinction.
    }
	\label{fig:AK}
\end{center}
\end{figure}

As has been shown in earlier systematic \HI\ ZoA surveys \citep[e.g.][]{Henning1997,hizoa2016}, the \HI\ detection-rate is completely independent of the foreground extinction at which the 2MASX galaxies are located. This is substantiated with Fig.~\ref{fig:AK} which shows the detections as a fraction (\%) of the total number of detections (N=230; red histogram) versus the $K$-band foreground extinction $A_{\rm K}$, together with the fraction over the total of the non-detections (N=773; blue histogram). The two histograms are indistinguishable. Non-detections do not occupy higher extinction levels than detections.

Figure~\ref{fig:AK} reveals furthermore that most of the observed 2MASX galaxies lie at intermediate extinction levels, i.e. 77.5\% lie at extinction levels below $A_{\rm K} < 0\fm5$, while 90\% have extinctions less than $A_{\rm K} = 0\fm7$ (i.e. $A_{\rm B} \sim 7\fm8$). Only 3.5\% of the bright 2MASX galaxy sample have higher extinctions than $A_{\rm K} =1\fm0$.

\subsection{Detections} \label{dets} 

The observed \HI\ parameters of the 220 clearly and 10 marginally detected target galaxies are listed in Table~\ref{tab:detsobsdata} as well as those of the 9 detected non-sample galaxies together with their most relevant near-infrared photometric measurements (observed values, uncorrected for extinction). An extract of the first ten galaxies is listed here. The complete table as well as all other tables are available online as supplementary material.

The columns in the table are as  follows:

\noindent {\sl Column 1:} 
2MASX J catalogue identifier for the J2000.0 coordinates used for the \HI\ observations. In case the galaxy is not listed in the 2MASX catalogue (Tables~A1c and~1d), the J2000-coordinates are given in brackets. Superscripts (flags) were appended to the name, where appropriate, to indicate the following:\\
\indent -- $\it a$ : further notes on this galaxy are given in the Appendix available as online supplementary material;\\
\indent -- $\it c$ or $\it c?$ : \HI\ spectrum (possibly) confused by one or more additional galaxies in the telescope beam; \\
\indent -- $\it d$ : derippled spectrum, that is, from which the effects of the baseline ripple were removed (see Sect. 3); \\ 
\indent --	$\it e$ : the target forms part of the higher extinction sample, i.e. $(|b| > 10\degr)$ \citep[cf.][]{Schroeder2018} and was observed for completeness; they are not used in the  ZoA data analysis; \\
\indent -- $s$ :  the target is not in the original, magnitude limited, sample but was
    observed to check on the detection of another target (in case the galaxy is
    not listed in the 2MASX catalogue, their coordinates are given in bracktes); \\
\indent --  $\it v$ : galaxy that has a published optical velocity inside the NRT search range; \\
\indent --  $\it v+$ : galaxy with an optical velocity which lies outside the NRT search range. 

\onecolumn
{\footnotesize
\input{Table1.tex} 
\smallskip
\input{Table2.tex}
}
\twocolumn

\noindent{\sl Column 2:} Other name.

\noindent{\sl Column 3:} The \K -band magnitude measured within the $r_{K20}$ isophotal aperture, {$K_{\mathrm 20}$}.

\noindent{\sl Columns 4 and 5:} The ($J$--$K$) and ($H$--$K$) colours measured within the $r_{{\rm K}{\mathrm 20}}$ isophotal aperture.

\noindent{\sl Column 6:} The elliptical aperture major axis at the 20 \masq\ \K -band isophote, {$d_{{\rm K}{\mathrm 20}}$} (in arcsec). 

\noindent{\sl Column 7:} The minor-to-major axial ratio $b/a$, fit to the 3-sigma super-co-added isophote. 

\noindent{\sl Columns 8 and 9:} The heliocentric \HI\ radial velocity determined as the centre of \HI\ profile measured at the 50\% peak flux density, \Vfifty, followed by its estimated uncertainty, $\sigma_{\rm V}$ (both in \kms). The latter was determined following the formalism in \citet{schneider86,schneider90},
  
  $\sigma_{v}=1.5(W_{20}-W_{50}){\rm SNR}^{-1}$, 
  
\noindent where \Wfifty\ and \Wtwenty\ are defined below, and $SNR$ is the signal-to-noise ratio (peak flux density divided by the rms).
  
\noindent{\sl Columns 10 and 11:} The \HI\ line widths measured at 50\% and 20\% of the \HI\ profile peak level \Wfifty\ and \Wtwenty , respectively, uncorrected for galaxy inclination (in \kms). Following Schneider et al., the uncertainty in the \Wfifty\ and \Wtwenty\ line widths is expected to be 2 and 3.1 times the uncertainty in \VHI , respectively.

\noindent{\sl Columns 12 and 13:} The total measured \HI\ line flux, \FHI, followed by its estimated uncertainty $\sigma_{\rm F}$ (in Jy \kms), 

$\sigma_{F}=2 (1.2W_{20} R)^{0.5}\sigma$,

\noindent where $R$ is the instrumental resolution in \kms (see Sec.~\ref{obs}); also derived according to \citet{schneider86,schneider90}.

\noindent{\sl Column 14:} The rms noise level values of the \HI\ spectra (in mJy). The rms noise level was determined over a 3000 \kms -wide part of the spectrum with low RFI occurrence. It is representative of the full velocity range as we have shown in a detailed analysis of about 50 spectra with little RFI and covering a wide range of observation time. There, we compared rms values in the velocity ranges $500-3500$ \kms, $4000-7000$ \kms\ and $7000-9000$ \kms\ and found no differences within the errors. 

\noindent{\sl Column 15:} The signal-to-noise ratio, S/N, determined by taking  the line width into account, following the ALFALFA \HI\ survey formulation from \citet{saintonge07}, i.e.

  S/N\,$= 1000(F_{\rm {HI}}/W_{50})\cdot(W_{50}/2 R)^{0.5})$, 
  
\noindent where $R$ is the instrumental resolution in \kms\ (see Sec.~\ref{obs}).

It should also be noted that the peak signal-to-noise ratio SNR (peak flux density divided by the rms) will be used in the text when discussing certain \HI\ profiles. This parameter should not be confused with the ALFALFA line-width-dependent signal-to-noise ratio, S/N (Column 15 of Table~\ref{tab:detsobsdata}.  Furthermore, the observed and derived \HI\ parameters of confused detections, contaminated by another galaxy in the telescope beam  (indicated by the $c$ flag in the Tables) are uncertain. Likewise, the fluxes of galaxies observed at an offset to their centre position may be underestimated (see the notes on individual galaxies in the online Appendix for details).

The \HI-line spectra of the detected galaxies are displayed as supplementary material (Fig.~A1) in order of Right Ascension. An example of the profiles of the ten first listed in Table~\ref{tab:detsobsdata} is given in Fig.~1. In each of the panels, the identifying 2MASX name (Column 1) is given at the top, and the rms of the respective NRT observation in the top left corner. The profiles have been smoothed to a resolution of 18~\kms. The full figure also contains the detections of non-target galaxies discussed below.\\

\begin{figure*}
\begin{center}
\includegraphics[width=.98\textwidth,angle=0]{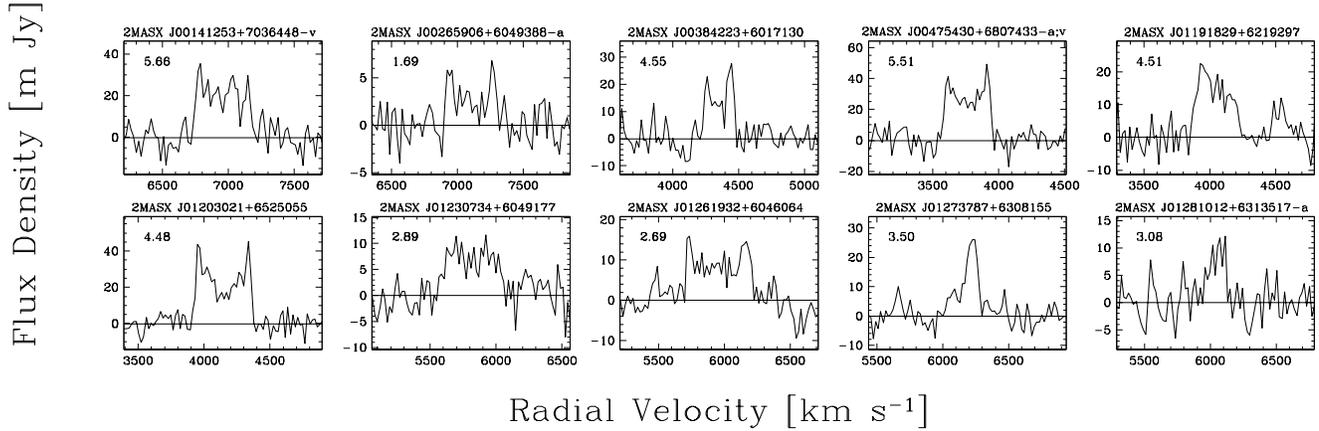}
	\caption{Example of the first two rows of the \nan\ 21cm \HI\ line spectra of detected galaxies. The velocity resolution is 18 \kms. The full figure of all detections (Table~1 and Table~2 combined) is available online as Fig.~A1. }
\label{fig:profile_example}
\end{center}
\end{figure*}

\noindent {\bf Detections of non-target galaxies:}
For completeness we list in Table~2 the cases where a galaxy other than the intended target was detected within the telescope beam -- both certain and marginal detections. The table is arranged in a similar manner as Table~\ref{tab:detsobsdata} but without the NIR parameters. An extract of Table~\ref{tab:nontargetdetsobsdata} is given on the preceding page; the full Table is available online, their spectra incorporated in the electronic-only Fig.~A1.

\noindent{\sl Column 1:} 2MASX name of the targeted galaxy. Although an \HI\ signal was identified, it has its origin in another galaxy. Where known the name of the source of the signal is given in the adjacent Column 2.
One further flag was introduced (as superscript), in addition to the ones mentioned under the description of Table~\ref{tab:detsobsdata}:\\
\indent --  $\it n$ : the coordinates mark the target position, not of the source of the observed emission, which is offset from the pointing position;
    
\noindent{\sl Column 2:} Name of the galaxy from which the \HI-signal in the beam of the target  originates (where identified). 

\noindent{\sl Column 3:} Distance between the pointing position and the actually detected, non-target galaxy in the Nan\c{c}ay beam (in units of 0.5 HPBW).

\noindent{\sl Column 4 -- 11:} The columns are identical to columns $8 - 15$ in Table~\ref{tab:detsobsdata}. All \HI -parameters are considered uncertain due to the off-beam detection. \\

\subsection{Comparison with other redshift measurements} \label{zcomp}

We started out with a target list of of bright 2MASX ZoA ($|b| < 10\degr$) galaxies without any prior redshift measurement in 2009. Since then, various redshifts for the original sample of 1003 galaxies have been observed. In addition, a renewed very careful literature search uncovered other measurements that had not been captured at the time the first target list was made available to us by Huchra. These are being used for a comparison here. We first discuss independent \HI-detections, followed by optical measurements in the Sect.~\ref{optcomp}.

\subsubsection{Comparison to other {\it H}\,{\it I} detections} \label{hicomp} 

For 42 of the 220 detected galaxies we found a total of 75 \HI\ measurements reported in the literature, many of these from the recent HIZoA survey \citep{hizoa2016}. Their \HI-parameters are given in Table~\ref{tab:HIdetscomp} together with the NRT parameters (the first ten lines are listed; the full table is available online). The entries in the table are:

{\footnotesize
\input{Table3.tex} 
}

\noindent {\sl Columns $1-5$}: provide the 2MASX name and main NRT \HI-parameters \Vfifty, \Wfifty, \Wtwenty and $F_{\rm HI}$ as listed in Table~\ref{tab:detsobsdata}.
 
\noindent {\sl Columns $6-9$}: gives the same \HI-parameters as reported in the literature.

\noindent {\sl Column 10:} gives the telescope code: 91m = NRAO 91m, ARE = Arecibo 305m, EFF = Effelsberg 100m, GBT = Green Bank Telescope, JBO = Jodrell Bank 76m, NRT = \nan\ 94m equiv., PKS = Parkes 64m, VLA = Very Large Array 27$\times$25m. 

\noindent {\sl Column 11:} The reference code {Ref.} refers to the paper which published the literature values. The references are given at the end of the table.

\begin{figure}  
\centering
\includegraphics[width=0.35\textwidth]{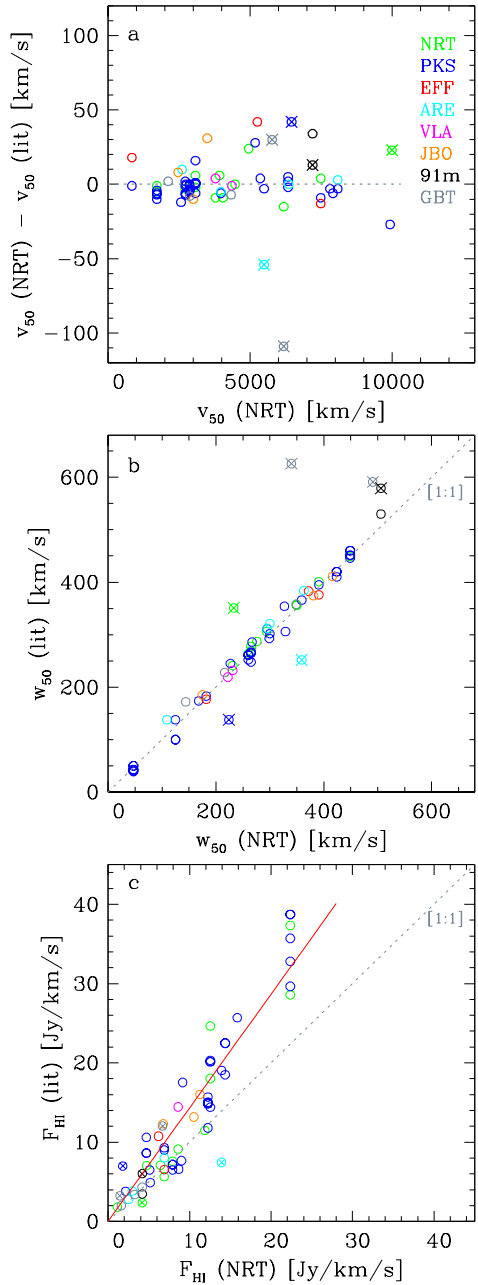}
\caption{NRT versus literature \HI\ line profile parameters. The colours identify the telescope used, the crosses (probably) confused profiles, the dashed lines equality, the red line (bottom panel) the slope (1.43) of a fit. 
Top panel:  Difference between the central line velocities as function of the NRT value. 
Middle panel: NRT \Wfifty\ values versus literature. 
Bottom panel: NRT \HI\ line fluxes versus literature.
}
\label{complitplot}
\end{figure}

For the following analysis, we exclude the reported detection of 2MASX~J05014040+4338109 by \citet{hauschildt87} because its \HI\ flux of 1.3$\pm$0.9 Jy~\kms\ is very uncertain. The spectrum shows a baseline ripple and the velocity of 5161~\kms\ agrees  neither with the NRT measurement of 7194~\kms\ nor with two other measurements of this galaxy. Figure~\ref{complitplot} presents the comparison of the central \HI\ velocities, \Wfifty\ line-widths and total line-fluxes. Six (probably) confused cases are marked with a cross and excluded from the analysis.

The central \HI\ velocities agree well, the mean difference and its standard deviation between our measurements and the literature values is $(-0.3\pm1.5$)~\kms\ and 1417~\kms, respectively. 

The \wfi\ line-widths also agree well: they have a mean offset of $(-3.7\pm1.6)$~\kms\ and a standard deviation of 12.1~\kms.
The six outliers (and likely confused cases) are\\
 -- 2MASX~J03480963+4955140: 233 vs 351~\kms, observed with the NRT, \\
 -- 2MASX~J04120068+3846073: 491 vs 591~\kms, observed with the GBT, \\
 -- 2MASX~J05014040+4338109: 506 vs 579~\kms, observed with the 91m at Green Bank, \\
 -- 2MASX~J05221454+3826469: 340 vs 626~\kms, observed with the GBT, \\
 -- 2MASX~J06225815+1108312: 359 vs 252~\kms, observed with Arecibo, \\
 -- 2MASX~J16463421-3903086: 224 vs 138~\kms, observed with Parkes.\\
Although their \wfi\ values are quite different, the \wtw\ widths for 2MASX~J06225815+1108312 and 2MASX~J16463421-3903086 are comparable (239 vs 260 \kms\ and 395 vs 336 \kms, respectively). 2MASX~J05014040+4338109 has no 91m \wtw\ width measurement as is the case for the two GBT detections. We find that 2MASX~J16463421-3903086 is confused in the Parkes beam, but is fine in our observation; 2MASXJ05014040+4338109 has a second 91m measurement which
agrees well with our observation, and 2MASX~J06225815+1108312 was also observed with Parkes and this agrees well with the NRT measurement. 2MASX~J03480963+4955140 was also observed with the NRT which suffers heavily from RFI above $v=8000$ \kms ; this could easily affect either of the the line width measurements. The difference in central velocity is acceptable though at 23 \kms .

The literature \HI\ line fluxes are on average 1.41$\pm$0.04 times higher than the NRT values (excluding the NRT confused target and non-target detections). A detailed comparison of \HI\ line fluxes measured recently at the NRT and at other telescopes (for both pointed and blind surveys) is presented in \citet{vandriel16} for the NRT NIBLES survey (see their Table A4). The NRT flux scale is consistent (0.91$\pm$0.08 here, 0.88$\pm$0.09 for NIBLES) with the single-horn Arecibo measurements of \citet{oneil04HI}. More than half (35/60) of the ZoA literature values we used for our flux scale comparison are from the Parkes HIPASS and HIZOA multi-beam surveys, for which we find a mean ratio of  1.44$\pm$0.04, which is consistent with the 1.34$\pm$0.28 found for NIBLES. 

The \HI\ flux scale comparison in \citet{vandriel16} focused on the differences between single beam pointed observations and total fluxes reconstituted from data cubes obtained for blind surveys, such as HIPASS at Parkes and ALFALFA at Arecibo. For the ALFALFA $\alpha$.40 catalogue \citep{haynes11} a mean ALFALFA/NRT flux scale ratio of 1.45$\pm$0.17 was found, similar to the HIPASS/NRT ratios mentioned above. Exploratory tests were made for NIBLES using an Arecibo multi-beam receiver data cube obtained for AGES (Minchin et al. 2010). These indicated that even for point sources changing the way in which total line fluxes are reconstructed can significantly change the result (see \citealt{vandriel16}). A full analysis is beyond the scope of NIBLES and the present work.

\subsubsection{Optical versus NRT redshifts}  \label{optcomp}
For 168 galaxies we have found 170 optical redshifts. We have compiled their values in Table~\ref{tab:optvel}, ordered by RA (again only the first 10 lines are listed; the full table can be found online). The columns provide the following information:
{\footnotesize
\input{Table4.tex}}

\noindent{\sl Column 1:} 2MASX identifier of the targeted galaxy. A further superscript $\it r$  was introduced to mark the galaxies with an optical velocity that fall in the velocity range affected by radar RFI ($8500-10,000$\,\kms).

\noindent {\sl Column 2:} The star under {$\Delta v$-flag} indicates whether the optical velocity differs from the $V_{\rm 50}$ NRT-values by more than 3 times the uncertainty in the optical velocity or by more than 150 \kms\ where no uncertainty was published.

\noindent {\sl Columns 3 and 4:} The heliocentric radial velocity measured in the optical {\vopt} and its estimated uncertainty, $\sigma_v$, (in \kms). 

\noindent {\sl Column 5:} The 'Ref.' codes  are detailed at the end of the table.

For eleven detected objects the difference between the \HI\ velocity and the optical velocity is larger than three times the uncertainty in the optical velocity, $\sigma_v$, whereas for four others without a published uncertainty it exceeds 150 \kms . Most (11/15) of these cases concern non-confused \HI\ detections. For eight of these, the velocity differences are quite significant, i.e., either more than $\sim$5$\sigma_v$ 
(2MASX J00475430+6807433, J02013241+6824219, J03292042+6601389, J04534877+4220445, J07483070-2532370, J16490239-3642570), or more than 300 \kms\ otherwise (J07403156-2618279 and J22540054+6728086); see also the notes in the online Appendix.\\

\noindent {\bf Non-target detections:} Of the clear and marginal non-target detections, seven and two  (28\% and 33\%) have published optical velocities. Not surprisingly, the majority (4 and 2 respectively)  have significantly different \HI\ velocities compared to the actual 2MASX target.

\subsection{Non-detections}  \label{nondets} 

As described earlier, the majority of the targeted galaxies were not detected in the 21 cm line. We present their positions, NIR parameters -- both observed and extinction-corrected -- as well as the rms of the NRT \HI-observation. The columns are a mixture of observed (as defined in Table \ref{tab:detsobsdata}) and extinction-corrected properties. The first ten galaxies are listed in Table~\ref{tab:nondetsobsprops}, the full complement is available online.
{\footnotesize
\input{Table5.tex} 
}

\noindent{\sl Columns 1 and 2:} 2MASX identifier and an alternate name (as in Table\ref{tab:detsobsdata}).

\noindent{\sl Columns 3 and 4:} Galactic coordinates (Column 3 and 4).

\noindent{\sl Column 5:} {$K_{\mathrm 20}$}-band magnitude measured at the $r_{K20}$ aperture (as in Table~\ref{tab:detsobsdata}). 

\noindent{\sl Column 6:} Galactic foreground extinction in the \K -band, estimated from the \citet{schlegel98}  DIRBE/IRAS extinction values $E(B-V)$ at the position of the source (henceforth SFD98). We correct the SFD98 $EBV$-values for the factor of $f=0.86$ derived in \citet{schlafly11}. Following \citet{Fitzpatrick99}, a conversion factor of $R{\rm_K} = 0.373$ was applied to arrive at the $A_{\rm K}$-value.

\noindent {\sl Column 7:} Extinction-corrected isophotal $K_s$-band magnitude, \Koc . The extinction correction follows the precepts of \citet{Riad2010}, i.e. the magnitude is not only corrected for the global foreground extinction at the position of the galaxy, $A_{\rm K}$, but includes an additional correction to substitute for the loss of the low surface-brightness (SB) features that have fallen below the $20^{\rm th}$ isophote because of the reduction in SB at each pixel of the galaxy's extent. Such a loss is not fully recovered when extrapolating towards magnitudes as has been shown conclusively in \citet[see their Fig.~1]{Said15TF}. We therefore prefer isophotal over total magnitudes when working in areas of extinction, or with low SB objects. We added the superscript $d$ (for diameter, respectively areal loss) to the standard superscript $o$ to emphasize this additional correction term to the standard $A_{\rm K}$ correction.

\noindent{\sl Column 9 and 10:} Extinction-corrected colours {$(J-K)\degr$, $(H--K)^{\circ}$} at the fixed $r_{\rm K20}$ isophote, hence corrected by ($A_J - A_K$) and ($A_H - A_K$) at said radius. Size corrections are superfluous for isophotal colours as long as the radius at which they are determined is fixed. 

\noindent{\sl Column 11:} The rms of the observation. The rms noise level was determined over a 3000~\kms -wide part of the spectrum with low RFI occurrence (as described in Table~\ref{tab:detsobsdata}). 

\begin{figure}
\begin{center}
\includegraphics[scale=0.85,angle=0]{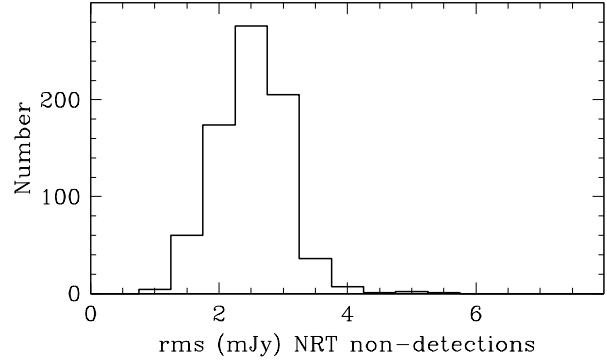}
	\caption{Distribution of the rms noise levels of the 771 \HI\ non-detections. Apart from a few exceptions our nominal completeness limit of 3.3mJy was reached throughout.
    }
	\label{fig:nondet}
\end{center}
\end{figure}

The distribution of the rms noise levels of all the NRT non-detections are displayed in Fig.~\ref{fig:nondet}. The  mean rms noise level of our non-detections is 2.5 mJy. This would imply a 3$\sigma$ detection limit of $\log($\MHI$) = 9.4$~\Msun, for a flat-topped profile with a \Wfifty\ of 300~\kms\ at a velocity of 5000~\kms, the mean values for our detections, hence nearly half a magnitude below the characteristic \HI-mass {$M^*_{\rm {HI}}$} \citep[e.g.][]{zwaan03,Martin+2010}.\\

\noindent {\bf Possible detections:} For 12 of the non-detection pointings, a tentative detection could be identified. None of these tentative detections has a published velocity to ascertain their reality. They do require higher sensitivity for confirmation. To not lose the information, nor confuse them with detections and marginals, these possible but unconfirmed detections are listed in a separate table (Table~\ref{tab:possdetsobsdata}) with their approximate \HI\ parameters. The columns are identical to Table~\ref{tab:nondetsobsprops}. 

\subsubsection{Optical redshifts of non-detections} \label{vopt_nondet}

Of the non-detections, 117 have published optical velocities, two of which with more than one determination. These are included in (Table~\ref{tab:optvel}) for information. In the following we use these measurements to assess whether we would have expected to detect this galaxy in \HI.

 In 24 cases, the optical velocity lies outside the NRT search range; their 2MASX names have been flagged with a ``$v+$''. Note however, that 2MASX J16490239-3642570 with \Vopt\ = 30,028~\kms\ was clearly detected by us at $\VHI= 6385$~\kms\ (Table~\ref{tab:detsobsdata}). 

A further 12 sources with published optical or \HI\ velocities in the range $8500-10,000$\,\kms\ are covered by radar RFI, which renders their detection nearly impossible. They have been added to Tables~\ref{tab:detsobsdata}, \ref{tab:HIdetscomp} and~\ref{tab:optvel}, but with the flag ``$r$''.  Despite the recurring RFI, we did detect seven galaxies in this RFI-plagued velocity range (Table~\ref{tab:nondetsobsprops}).

Four of our non-detections have \HI\ detections reported in the literature (added at the end of Table~\ref{tab:HIdetscomp}): \\
-- 2MASX J02121002+6144326 (\VHI\ = 5900 \kms, \citealt{Kerr1987}),\\ 
-- 2MASX J07003437-1020151 (\VHI\ = 9608 \kms, \citealt{hizoa2016}),\\ 
-- 2MASX J20401346+5059165 (\VHI\ = 2783 \kms\ and \Vopt\ = 3028 \kms,
\citealt{paturel03,huchra12}), \\  
-- 2MASX J22131198+6153077 (\VHI\ = 3839 \kms, \citealt{seeberger94}).\\

It is not surprising that we did not confirm these detections. The first one has a positional uncertainty of $5\arcmin - 10\arcmin$ which means it could be a hidden galaxy outside of our NRT beam; its reported line flux could not be reduced to a value in Jy \kms\ (see online Appendix). The second one is covered by radar RFI at the NRT and the third is a very weak detection  (the reported peak SNR would be only $1.6 \times\ $~rms). For the fourth galaxy,  the reported peak SNR and its uncertainty corresponds to $4.7\pm1.3 \times$~rms noise level -- this marginal-looking detection we could, however, not confirm.

\section{Observed and global properties of the detected galaxies} \label{properties}

When discussing and interpreting the parameter space of the newly obtained NRT \HI-data, caution is advised. This second paper of our bright 2MASX ZoA galaxy series is aimed at completing -- as best as possible -- the existing redshift gap in the zone of obscuration using the 21cm line emission of neutral gas which is not affected by dust obscuration. The observed sample therefore does {\sl not} constitute a complete sample, despite the imposed NIR magnitude completeness limit, because only galaxies which at the start of our observing campaign had no prior redshift were targeted (see Sect.~\ref{sample}). Of course, the large fraction of non-detections also play a role, but that is a bias we have in common with all targeted \HI-observations of galaxies. Our  sample is close to a targeted magnitude complete sample for the inner ZoA $(|b| < 5\degr)$, for which only a small fraction had published redshifts.

\subsection{Observed \HI-parameters}

\begin{figure}
\begin{center}
\includegraphics[scale=0.75,angle=0]{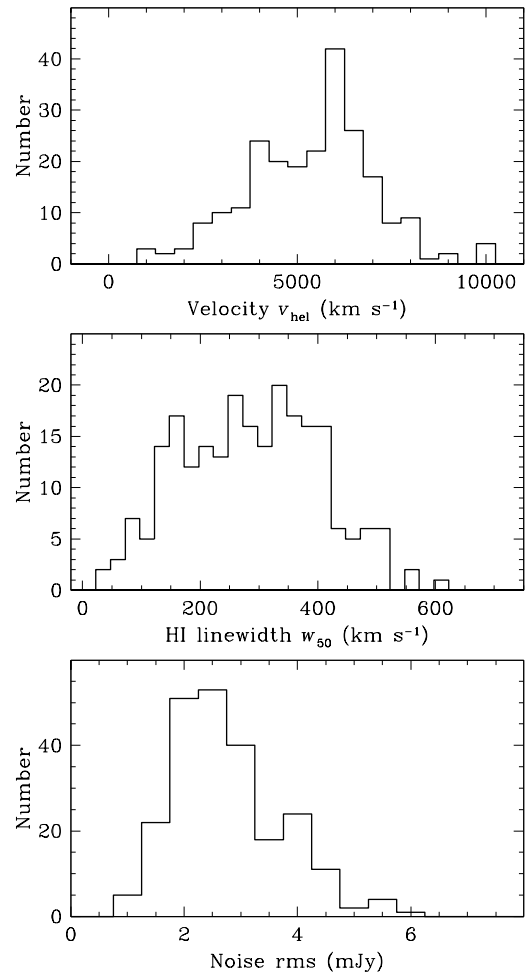} \label{hi-params}
	\caption{\HI-parameters of the detected galaxies ($N=230$) and include, from top to bottom: the heliocentric velocity, the \wfi\ line width, and the rms noise histograms. All panels include the 10 marginal detections (Table~A1).
    }
	\label{fig:hiparams}
\end{center}
\end{figure}

In Fig.~\ref{fig:hiparams} we present histograms of the \HI-parameters of the detected galaxies including the 10 marginals.  The different panels represent the velocity distribution, \Vfifty, the line width measured at the 50\%-level, \Wfifty, and the rms of the detected galaxies, and  will be discussed in this order. 

The velocity distribution in the top panel is mostly representative of the inner ZoA ($|b| < 5\degr$) since 180 out of the 230 detections (78\%) lie within that latitude range. A further 44 galaxies (20\%) were found in the latitude strips $5\degr < |b| < 10\degr$ (see also Fig.~\ref{fig:sample}). The other 5 detections (2\%) were made at higher latitude in the  high-extinction pockets. 

Galaxies are found over most of the observed redshift range, with the majority lying within $2000-8000$~\kms. The rather strong fall-off for $V \ga 8000$~\kms\ is mostly due to recurring radar RFI around 8500~\kms,  particularly notable in Fig.~\ref{fig:sense}, in addition to the loss of sensitivity given our rms limit of 3~mJy.  The counts pick up again around 10\,000~\kms\ where we find four high \HI-mass detections. Although not quite as persistent, RFI may have affected the detection rate around $\sim 3500~\kms$.

The velocity histogram shows two clear peaks over and above what is expected from a more homogeneous galaxy distribution, a minor peak around 4000~\kms\ and quite a prominent one with broad shoulders at 6000~\kms. As will be seen in the large-scale structure section, both of these peaks are associated with the Perseus-Pisces Supercluster (PPS) through two different entry points into the ZoA, as well as a completely new and unsuspected filament/wall-like structure in between those two.

\onecolumn
{\footnotesize
\input{Table6.tex}
\bigskip
\input{Table7.tex}
}
\twocolumn

The line width distribution (second panel) has quite an interesting shape in the sense that its distribution is nearly flat between line width of $150 <  \Wfifty < 400$~\kms, with few narrow and also broad line-widths spirals. The latter is more standard. This is very different to either systematic 'blind' \HI-surveys or targeted \HI\ follow-up surveys.  For instance, the line width distribution of the HIZoA \citep{hizoa2016}  with a rms of 6~mJy finds the large majority of detections between \Wfifty\ of $50 - 200\kms$, followed by a rapid decline for increasing line width above 200~\kms\ (see their Fig.~2). The NRT detection sample is probably more directly comparable to the Parkes \HI-observations of optical ZoA galaxies pursued by \citet{2002A&A...391..887K, Schroder2009} which has a similar rms limit and also primarily targeted galaxies without  prior redshift measurement. Although not presented in these papers, the \Wfifty\ distributions are not flat, however, but rather Gaussian in shape with a peak around 220~\kms\, followed by a slow decline towards the higher values. While the underrepresentation of very low line-width galaxies in our NRT sample can partly be explained by the NIR-selection which inherently biases against blueish, low surface-brightness gas-rich dwarfs, it does not explain the overall flatness of the \Wfifty-distribution.

The last panel shows the rms of the detections. The mean is a bit higher than that of the non-detections, 2.78 versus 2.50~mJy (Fig.~\ref{fig:nondet}), while its dispersion is broader (0.96 vs 0.54~mJy). It has a higher number of detections with lower rms values ($\la 3$~mJy) which is due to the extra integration that was added to firm up weaker detections. There also is a higher number towards larger rms values. This  is due to our survey strategy of starting with shorter integrations to capture high signal-to-noise profiles early on, and not waste integration time.

\subsection{Global properties}

In Table~\ref{tab:detsderprops} the global parameters of the 230 \HI\ detected galaxies, including NIR properties such as extinction-corrected magnitudes and colors, and \HI-mass to light ratios are listed. All global properties are based on a Hubble constant of $H_{\mathrm 0} = 75~$\kmsMpc. The 10 marginal detections of targeted 2MASX ZoA galaxies are added at the bottom of the table. The full table is online, the first 10 galaxies are shown in the sample table presented here
. The explanation of the columns are:

\noindent {\sl Column 1:} Same as column 1 in Table 1.

\noindent {\sl Columns 2 and 3:}  Galactic longitude and latitude, $\ell$ and $b$, of the target galaxy.

\noindent {\sl Column 4:} Galactic foreground extinction $A_{\rm K}$ estimated from the \citet{schlegel98} (SFD98) DIRBE/IRAS extinction values $E(B-V)$ at the position of the source (see Table~\ref{tab:nondetsobsprops} for further details). 

\noindent {\sl Column 5:} Distance $D = V_{\rm {LG}}/H_{\mathrm 0}$, where the target's radial velocity (in \kms) with respect to the Local Group was adopted as $V_{\rm {LG}}=V_{\rm {HI}}+300\sin{l}\cos{b}$, with a Hubble constant of $H_{\mathrm 0} = 75~$ \kmsMpc.

\noindent {\sl Column 6:} 
Extinction-corrected isophotal $K_s$-band magnitude, \Koc\ measured at the \rkc\ aperture. The extinction correction follows the precepts of \citet{Riad2010}. 

\noindent {\sl Column 7 and 8:}  Extinction-corrected colours {($H$--$K$)$\degr$, ($J$--$K$)$^{\circ}$} determined within the $r_{{\rm K}{\mathrm 20}}$ isophotal radius (in mag). No additional size correction has been applied since isophotal colours are always measured at fixed aperture/radius.

\noindent {\sl Column 9:} Radius \rkc\ determined at the extinction-corrected surface brightness level of 20 \masq\ in the \K -band (in kpc), hence corrected for the loss of the outer areas that lie below the $20^{\rm th}$ isophote due to extinction (see also columns 6 and 7 in Table~\ref{tab:nondetsobsprops}).

\noindent {\sl Column 10:} Extinction-corrected \K -band luminosity {\LKc:}   within $r^{\rm o,d}_{\rm K20}$ (in \Lsun), for an assumed solar absolute magnitude of $3\fm31$ \citep{colina97} and the galaxy's  distance  $D$.  

\noindent {\sl Column 11:} Total \HI\ mass {\MHI}, where {\MHI}$ = 2.356 \times 10^5 \cdot D^2 \cdot F_{\rm HI}$ (in \Msun). 

\noindent {\sl Column 12:} {\MHILKc:} ratio of the total \HI\ mass to the extinction-corrected \K -band luminosity (in solar units).

\noindent {\sl Column 13:} {\vrot:} rotation speed corrected for inclination $i$; $v_{\rm {rot}}=W_{50}/2\sin(i)$ for
  $\sin(i)<0.2$, whereas for galaxies with higher inclinations we assumed 
	$v_{\rm {rot}}$ = \Wfifty/2 (in \kms).

\noindent {\sl Column 14:} {\Mdyn:} dynamical mass, estimated as 
$M_{\rm dyn} = v_{\rm rot}^2 \cdot r^{\rm o,d}_{{\rm K20}}/G$ (in \Msun).

\noindent {\sl Column 15:} {\Mbardyn:} the ratio of the combined \HI\ and stellar baryonic mass as a fraction of the total dynamical mass, where \Mbar\,$ = 0.8 L_{\rm K} + 1.4$\MHI\ \citep{mcgaugh00}.

\begin{figure}
\begin{center}
\includegraphics[scale=0.9,angle=0]{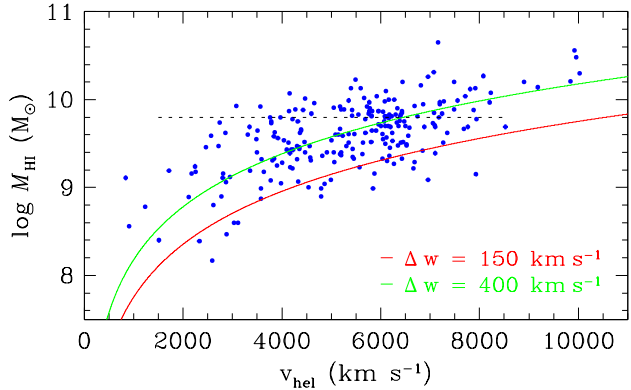} \label{MHI_sens}
	\caption{\HI-mass versus velocity plot for the 230 detected galaxies. The two displayed $3~\sigma$ sensitivity curves represent line widths of $\Delta W_{50} = 150\,$\kms\ and 400\,\kms, the range covered by the majority of our detections (see panel 2 in Fig.~\ref{fig:sense}). The dashed black line is indicative of log~\MHIstar.
    }
	\label{fig:sense}
\end{center}
\end{figure}

We now look at some of the global properties of the detected galaxies. Figure~\ref{fig:sense} displays the \HI\ mass versus recessional velocity. The coloured curves indicate the 3-$\sigma$ \HI-mass detection limits for the NRT sensitivity limit of rms~$ \sim 3$\,mJy and line widths of $\Wfifty = 150~\kms$ and 400~\kms\ respectively, the range that encompasses the majority of the detected galaxies.
The figure reveals some clustering that coincides with the peaks in the velocity histogram (top panel of Fig.~\ref{fig:hiparams}) and confirms that most of the detected galaxies hover around the PPS redshift of $cz \sim 6000$~\kms. At this distance the \HI-masses of our detections lie between log~\MHI\ of $9 - 10.5$~\Msun. Comparing this to the $3~\sigma$ sensitivity curves indicates that we sample well below the characteristic log \HI-mass of  9.8~\Msun\ \citep{Zwaan+2005}, marked in Fig.~\ref{fig:sense} as a dashed black line. 

\begin{figure}
\begin{center}
\includegraphics[scale=0.8,angle=0]{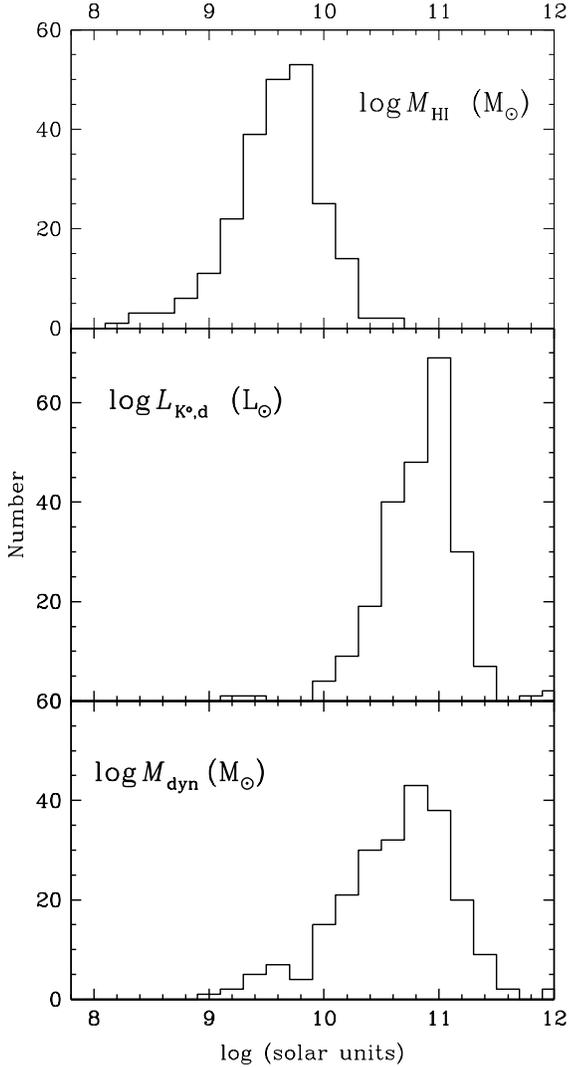}
	\caption{Global parameters of the certain and marginally detected galaxies, displayed as histograms. These are the logarithms of the HI-mass \MHI, the extinction-corrected \K-band luminosity $L_{\rm K}$, and the dynamical mass \Mdyn\ (from top to bottom).
    }
	\label{fig:glob}
\end{center}
\end{figure}

In Fig.~\ref{fig:glob} we present the \HI\ mass, the extinction-corrected \K-band luminosity $L{_{K}^{d,o}}$, and the dynamical mass \Mdyn. The \HI-masses of the NRT detections range from \lMHI$ = 8.2 - 10.6$~\Msun\ with the majority (80\%) confined to the narrow range of $9 - 10$~\Msun. The distribution is fairly symmetric around its mean (9.6~\Msun). Its slightly offset maximum coincides closely with the characteristic \HI-mass, log~\MHIstar$\,= 9.8$~\Msun\ derived by \citet{Zwaan+2005} for the HIPASS survey \citep{Meyer2004}. The histogram confirms that we are not sensitive to low \HI-mass galaxies, which are more prevalent in systematic \HI-surveys like HIZoA, HIPASS and ALFALFA. Our detection rate is also a bit low on the high-mass range. The reason is  that many of these will have been identified already in the shallower (rms $\sim 6$~mJy) 'blind' HIZoA.

The middle panel displays the respective luminosities of these same galaxies. The distribution is quite similar in shape as that of the \HI-masses, albeit shifted in log space by a value of $\Delta = 1.2$  to log\,$L_{\rm K} = 10.8$\,\Lsun, with a slightly lower dispersion. It reveals a similar tail towards lower values. The similarity of the log-distributions is reflected in the average \HI-mass-to-light ratio which for this sample is $\MHILK = -1.23$\,\Msun$/$\Lsun, in a narrow Gaussian distribution with a $\sigma$ of 0.33\,\Msun$/$\Lsun.

The bottom panel presents the dynamical mass \lMdyn. The peak as well as the mean of the distribution are close to that in the luminosity distribution, shifted only by $\Delta = 0.2$ towards higher luminosities in log-space. The shape itself bears closer similarity to the \HI-mass distribution, apart from the shift of 1.4. It is therefore not surprising that the ratio of the combined \HI\ and stellar baryonic mass as a fraction of the total dynamical mass, is quite low on average, i.e. \Mbardyn$ = 0.17$, with a dispersion of $\sigma = 041$.

\begin{figure*}
\begin{center}
\includegraphics[width=.66\textwidth,angle=270]{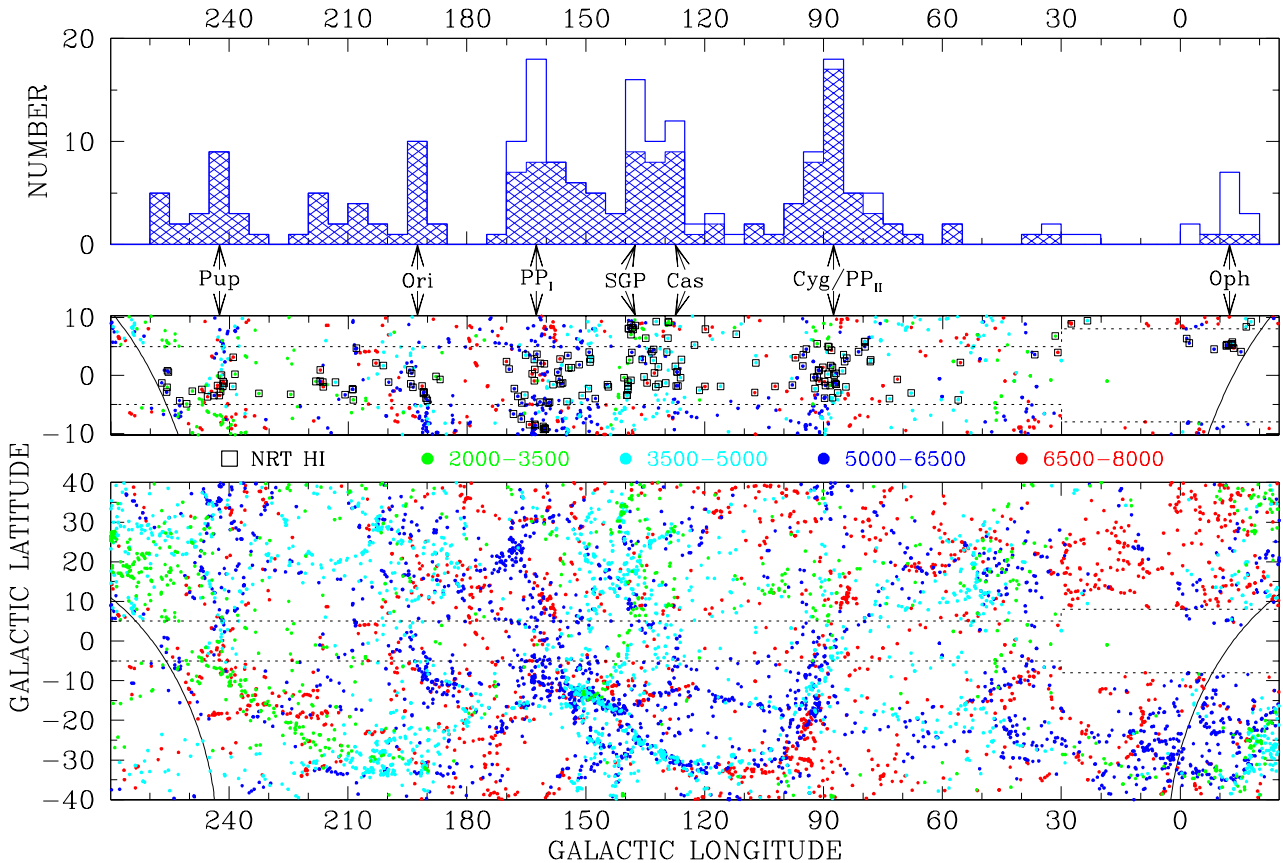}
	\caption{Top panel: Histogram of the NRT \HI-detections within $|b| < 10\degr$ as a function of Galactic longitude, where hashed reflects the inner ZoA ($|b| < 5\degr$) with prominent peaks being labeled according to constellation. Bottom two panels: On-sky distribution in Galactic coordinates of the NRT detected galaxies in the ZoA, together with 2MRS-galaxies down to the same extinction-corrected NIR magnitude limit ($K^o_s \le 11\fm25$). Galaxies are displayed for the velocity range $2000-8000$\,\kms\ in which most (94\%) of the NRT detections reside. Colours indicate  redshift intervals of 1500\,km/s width. The middle panel displays the survey area with the new detections outlined by squares. The bottom panel displays a wider area for context of the large structures. The dashed lines mark the 2MRS exclusion zone ($|b| < 5\degr$; respectively $|b| < 8\degr$  for $0\degr \pm 30\degr$). The black lines mark the southern NRT declination limit.  
   }
	\label{fig:onsky}
\end{center}
\end{figure*}

All these panels show a similar distribution to what was found -- with much smaller numbers -- for the NRT detections in our pilot project \citep{vandriel09}.

\section{Unveiled Large-Scale Structures} \label{lss} 

The description of the uncovered large-scale structures (LSS) revealed by the NRT detections is described below in context to known structures adjacent to the most obscured part of the Milky Way. Some preliminary results were given in \citet{ramatsoku12} and \citet{ramatsoku14}. 
 
Figure~\ref{fig:onsky} displays in the top panel a histogram of the NRT detections as a function of Galactic longitude $\ell$ for the $\Delta \ell$-range visible from \nan. The hashed part represents the new detections within $b < \pm 5\degr$, while the open histogram includes all the NRT detections within $b < \pm 10\degr$ (but not the higher latitude EBV sample). 
The histogram identifies the newly unveiled filamentary ZoA crossings quite clearly. We have labeled the various peaks according to constellation (Puppis, Orion, Cassiopeia, Cygnus and Ophiuchus) or features associated with known large-scale structures (the Perseus-Pisces supercluster PP${\rm_I}$, the Supergalactic Plane SGP, and the second crossing that seems to form part of PPS crossing in Cygnus, dubbed PP$_{\rm II}$). We will use this terminology henceforth for discussions of the structures. The highest peaks in the histogram are located within the previously hardly explored longitude range $70\degr < \ell < 180\degr$. As described in Sect.~\ref{intro}, this range is bounded by ZoA longitude strips accessible to Arecibo ($\ell \sim 180\degr - 215\degr$ and $\ell \sim 70\degr - 35\degr$), with some \HI-detections on either end of these strips from HIZoA and its northern extension \citep[$\ell > 196\degr$, respectively $\ell < 52\degr$;][]{hizoa2016, Donley2005}.

The middle and bottom panel of Fig.~\ref{fig:onsky} present the NRT detections in on-sky plots {\sl merged} with redshifts available in the literature for 2MASX galaxies with the same  magnitude limit ($K_{\rm s}^{\rm o} = 11\fm25$). They hence constitute a low-latitude complement of the first 2MRS data release and the 2MTF survey. The galaxies are colour-coded in redshift shells of widths $\Delta V = 1500$\,\kms, as a visual aid when tracing the continuity of structures in the redshift dimension. The galaxies below 2000\,\kms\ were not plotted to not dilute the traced structures. Moreover, only five galaxies (2\%) were detected at these low redshifts. This is also true for the high redshift range ($>8000$\,\kms) where the scarcity of redshifts (4\%) due to survey sensitivity and RFI would not add much to the discussion.

The middle panel focuses on the 2MZOAG survey region ($|b| < 10\degr$). Here the NRT detections are outlined by squares. The bottom  extends to latitudes of $|b| = 40\degr$ but without the squares to provide an unhampered view of how the ZoA structures link to the cosmic web beyond the ZoA. Note again the predominance of NRT detections in the $180\degr < \ell < 70\degr$ range: hardly any redshifts were known in the inner ZoA before the systematic 2MZOAG \HI-follow-up observations. 

In the following the new structures will be described in more detail. Note at the same time the redshift slices out to $v_{\rm hel} < 12\,000$\, \kms\ that are presented in Fig.~\ref{fig:pizza} for the innermost ZoA ($|b| < 5\degr$, top panel) or the full range of the 2MZOAG survey ($|b| < 10\degr$, bottom panel). In the redshift wedges, red dots mark the new detections, while the blue crosses are 2MASX galaxies as listed in the 2MZOAG catalog. 
\smallskip

\noindent {\bf Puppis:} The filamentary crossing around $\ell \sim 245\degr$ has first been explored by \citet{kraan-korteweg92}. \HI-follow-up observations of optically visible galaxies with the 100\,m Effelsberg telescope uncovered various nearby groups at $ v\sim 1500$ and 3000\,\kms. These have been corroborated by \citet{hizoa2016} with further indications of a condensation at $v \sim 7000$\,\kms. The NRT results add further detections to these groupings, enhancing in particular the galaxy agglomeration around 7000\,\kms\ (see also Fig.~\ref{fig:pizza}). 
\smallskip

\noindent{\bf Orion:} A notable peak in the histogram is found a bit further on around $\ell \sim 190\degr$ -- located around the boundary of the Orion constellation (its top part) and Gemini (lower part), with a typical redshift around $v \sim 5000$\,\kms. This crossing forms part of the extension of the so-called Gemini-Monoceros filament \citep{takata94} which is weakly visible in the bottom panel of Fig.~\ref{fig:pizza}. It seems to extend from ($\ell,v) \sim (200\degr,4000\,\kms)$ to $(190\degr,6000\,\kms)$.  The low velocity end of the Monoceros filament ($\ell \sim 210\degr$) discussed in the HIZoA surveys \citep{hizoa2016,Donley2005} is, however, not as prominent in the 2MASX redshift distribution. 

The filament might further be connected to the N1600 supercluster \citep{saunders1991} at ($\ell,b,v) = (194\degr, +24\degr, 4400\,\kms$). Indeed, the bottom panel of Fig.~\ref{fig:onsky} shows this filamentary crossing connecting to N1600 and back into the Plane around $\ell \sim 165\degr$ (discussed below) seems to be part of a circle surrounding a void centered at $\ell \sim 180\degr$. Given its velocity range (see the dominance of cyan and blue dots) it is not inconceivable that this structure is somehow interlinked with the PPS wall crossing. 
\smallskip

\noindent{\bf Perseus-Pisces (PP$_{\rm I}$) and its extension to Cygnus (PP$_{\rm II}$): }\\
\noindent{\sl The hidden cluster 3C\,129:} One of the highest peaks in the velocity histogram is centered around $\ell \sim 160\degr$. It looks less well-defined in the bottom panel of Fig.~\ref{fig:onsky}, but extends nearly vertically across the full width of the ZoA. The galaxies pertaining to this structure lie in the velocity range $v \sim 6000 \pm 1000$\,\kms. Interestingly, two very strong radio galaxies are found along the path of this broad feature. Their position and redshifts \citep[$v = 6236$ and 6655~\kms, respectively;][]{spinrad75} confirm that they reside inside the galaxy concentration. These are the head-tail radio source 3C\,129 and the double-lobed giant elliptical radio galaxy 3C\,129.1. The presence of radio sources with bent lobe morphology usually is indicative of a rich cluster environment. Indeed, hot X-ray emission surrounding these two radio galaxies was observed as part of the CIZA (X-ray Clusters in the ZoA) survey, i.e. CIZA~J0450.0+4501  \citep{ebeling00}. The hot X-ray gas is centered on the radio galaxy 3C\,129.1 which lies deep in the ZoA ($b = 0\fdg27$). We do not see a prominent finger-of-God in Fig.~\ref{fig:pizza} at the position of this cluster. In fact, hardly any NRT detections 
were made at the cluster core; this may well be caused by gas removal processes associated with the observed hot cluster gas of CIZA~J0450.0+4501. In addition, NIR observations also suggested this cluster to be quite rich: despite being mostly invisible in the optical it stands out as a beacon in the photometric redshift sky distribution of the 'full' 2MASX catalogue \citep[][e.g. Fig.~1]{Jarrett2004}. 


To put this new cluster and wall-crossing into context to the known structures adjacent to the ZoA, we list the most prominent clusters that define the quite distinct chain-like wall of the PPS (see cyan and blue dots below the Plane for the longitude range $\ell \sim 120\degr$ to 160\degr, in the bottom panel of Fig.~\ref{fig:onsky}, in order of {\sl descending} latitude. In this list we have already included Abell~569, which lies above the Galactic Plane, and the new ZoA link given by the cluster 3C\,129 (or CIZA~J0450.0+4501) at the Galactic equator.
The clusters of the main PP complex then are:\\
\indent Abell~569: \, \, $(\ell,b,v) \sim (168\degr,+23\degr, 6000\,\kms$)\\
\indent 3C~129:  \,\, \, \, \, $(\ell,b,v) \sim (160\degr,+\,.3\degr, 6300\,\kms$)\\
\indent LDCE~293\footnote{\citet{crook08} group catalog based on the  first 2MRS data release.}:  $(\ell,b,v) \sim (162\degr,-10\degr, 6000\,\kms$)\\
\indent Perseus Cl:\, \, $(\ell,b,v) \sim (151\degr,-13\degr, 5400\,\kms$)\\
\indent Abell~347: \, \, $(\ell,b,v) \sim (141\degr,-18\degr, 5500\,\kms$)\\
\indent Abell~262: \, \,  $(\ell,b,v) \sim (137\degr,-25\degr, 5200\,\kms$)\\
\indent Pisces~Cl: \,\,\, \,  $(\ell,b,v) \sim (127\degr,-32\degr, 5400\,\kms$)\\
Note that the Pisces Cloud really consist of a chain of three prominent galaxy groupings which from left to right are the NGC~507, NGC~383, and NGC~266 groups at the mean redshifts of 5700, 5200, and 4900\,\kms\ respectively.

\begin{figure}
\includegraphics[width=.48\textwidth,angle=0]{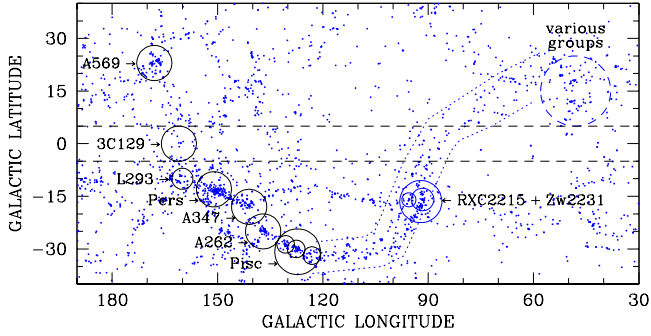}
	\caption{The Perseus-Perseus supercluster and its extensions across the ZoA. The figure is an extract of Fig.~\ref{fig:onsky} limited to the PPS-velocity range (4500--7000\,\kms) range. The various clusters and groups that form the primary chain (black circles) and its continuity and re-emergence into the ZoA are indicated (blue circles and dashed lines).}
    \label{fig:PPS}
\end{figure}

\begin{figure}
\begin{center}
\includegraphics[width=.5\textwidth]{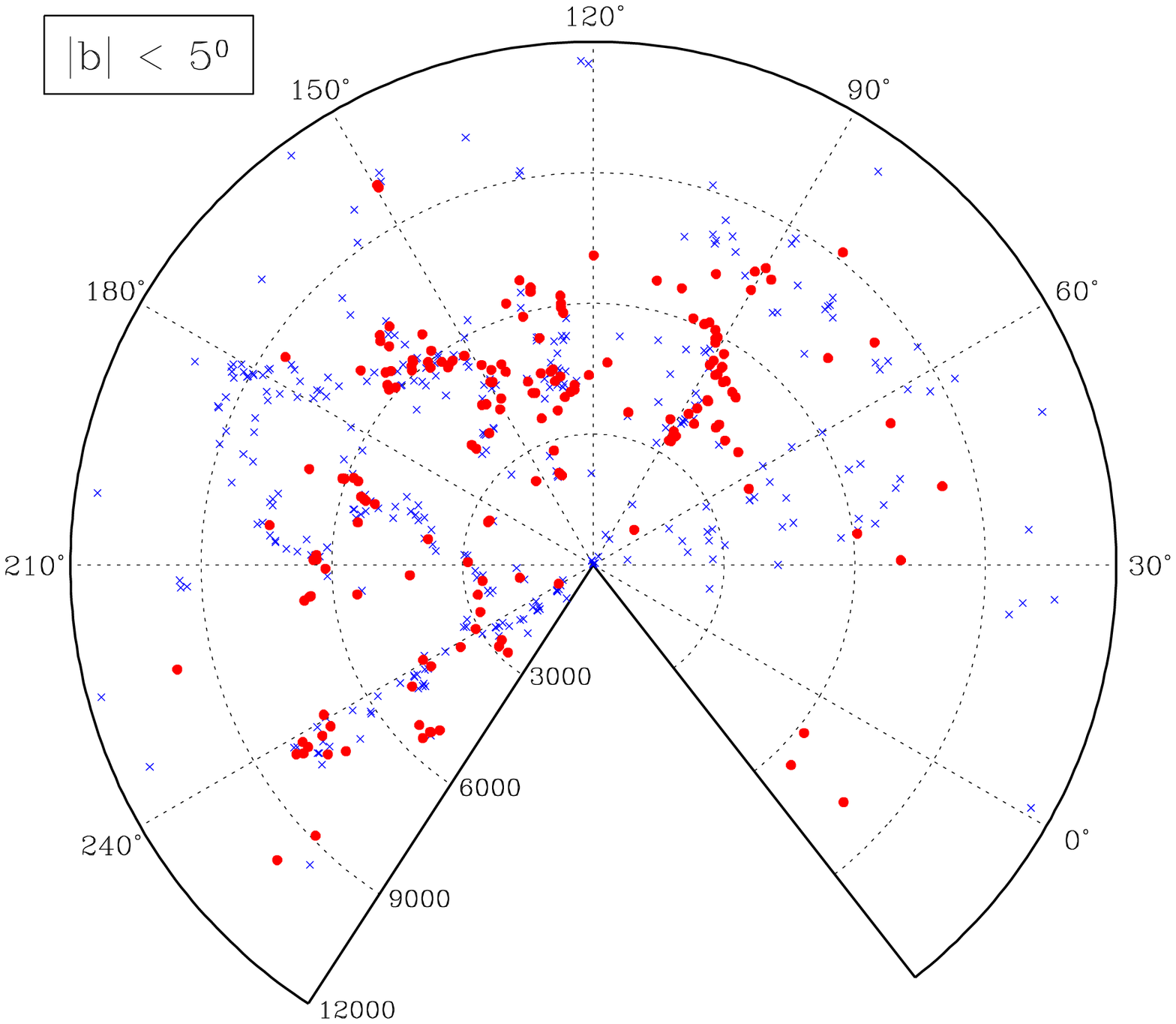}\\
\vskip 0.5cm
\includegraphics[width=.5\textwidth]{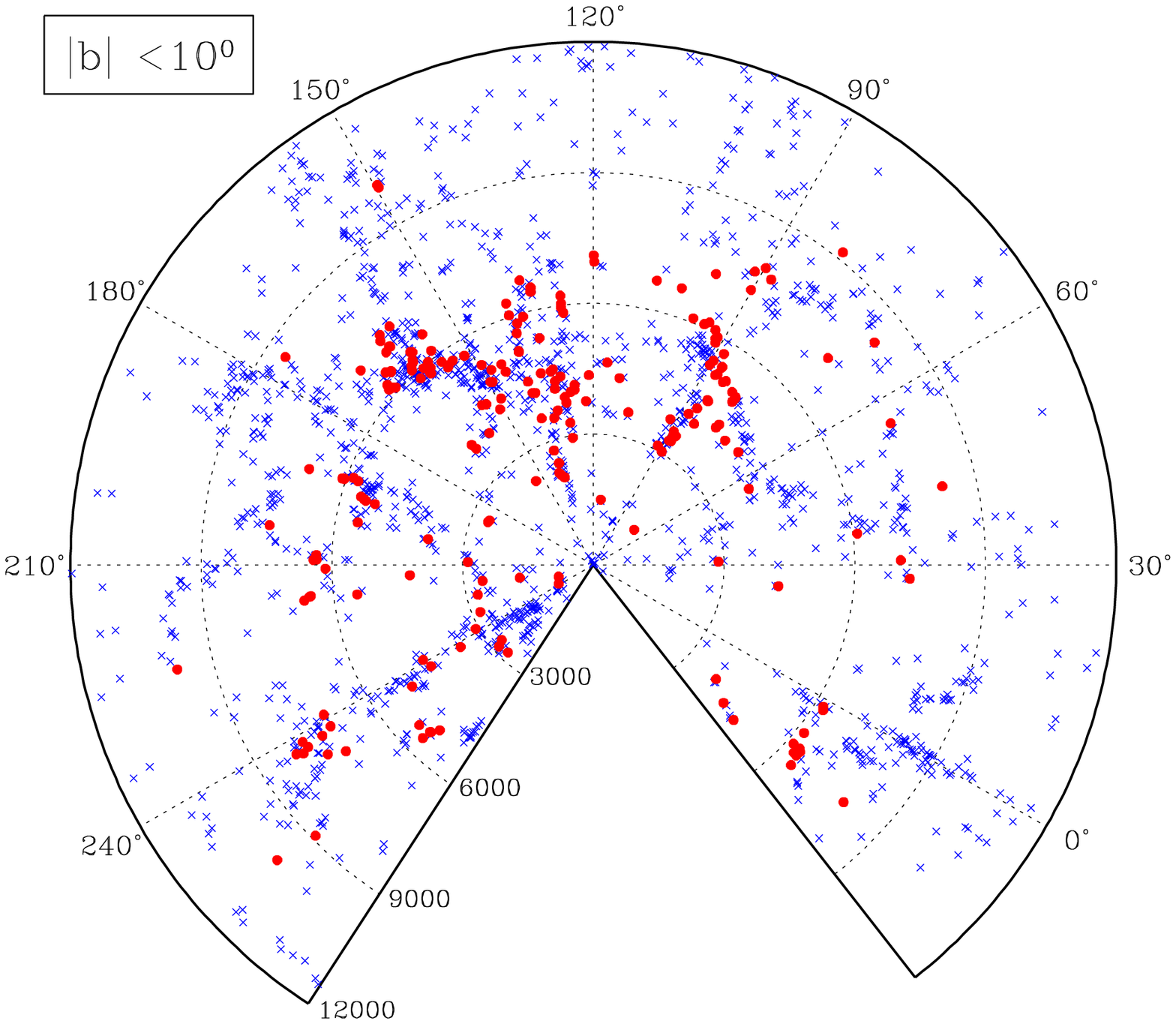}
\caption{Redshift wedge out to $V_{\rm hel} < 12\,000$\,\kms\ centred on the longitude range of the \nan\ 2MZOAG survey. Red dots represent NRT \HI-detections, and blue crosses 2MASX galaxies with prior redshift measurements. The top panel displays the inner ZoA ($|b| < 5\degr$) and the bottom panel the full width of the 2MZOAG survey ($|b| < 10\degr$). The most prominent new ZoA feature is the wall-like structure that forms part of the PPS (running from about $\ell \sim 125\degr$ at 4500\,\kms\ to 165\degr\ and 6500\,\kms. The latter includes the new filament around $\ell \sim 165\degr$ in which the 3C\,129 cluster is embedded.
   }
	\label{fig:pizza}
\end{center}
\end{figure}

For a better visualisation of the PPS complex, we present with Fig.~\ref{fig:PPS} an additional on-sky illustration. It is similar to the bottom panel of Fig.~\ref{fig:onsky} but dedicated solely to the longitude and distance range which bears relevance to uncovered ZoA crossings PP$_{\rm I}$ and PP$_{\rm II}$) in Perseus and Cygnus. It is therefore restricted to the redshift range of $4500-7000$\,\kms\ and longitudes $30\degr - 190\degr$. The above listed clusters are identified in this plot as black circles with shortened labels corresponding to the above list, while the extension of the PP-chain towards Cygnus, and galaxy agglomerations embedded within it, are indicated with a blue trace.

The 3C~129 cluster and the wall in which it is embedded is located exactly in the extension between A~569 and the Perseus cluster (A~426). The herewith extended composition of the PP chain suggests the supercluster to be at least $\Delta \ell \sim 70\degr$ long at the mean distance of 5000-6000\,\kms. The concept of a continuation of the PPS across the ZoA towards Abell~569 in the northern Galactic hemisphere is not new. It was already hypothesised by \citet{gregory1981}. Concerted efforts were made in the past to trace this prospective connection \citep{focardi84, focardi86, Chamaraux1990, lu95, pantoja97, saurer1997} through \HI-follow-up observations of optically identified galaxies. While the results were generally supportive of a connection, the data in this dust-enshrouded region were too sparse to be conclusive, partly because the hidden ZoA PP cluster discussed below remained inconspicuous. 

Once the relevance of the optically hidden 3C~129 cluster as a new constituent of the PPS became apparent in the evolving NRT survey, a deep follow-up \HI\ imaging survey project was launched 
with the Westerbork Synthesis Radio Telescope (WSRT) of a mosaic of 35 WSRT fields  (9.6\,sq\,deg), which includes the cluster and part of the wall in which it is embedded. Details can be found in \citep{ramatsoku16}. Suffice to say that the cluster is confirmed as a rich cluster, showing evidence of \HI-deficiency at its core. Moreover, the main 3C\,129 cluster seems to be undergoing a merger with a galaxy concentration that is in-falling along the PPS wall.
\smallskip

\noindent{\sl The Perseus-Pisces extension (PP$_{\rm II})$ in Cygnus: }
The PPS chain does not abruptly end with the Pisces cluster: Fig.~\ref{fig:PPS} shows that the chain seems to continue further on from the Pisces Cloud, albeit with a slightly less populated, thinner filament composed of an alignment of smallish groups, to about $(\ell,b) \sim~(110\degr,-30\degr)$,  marked by a blue dotted line, from where it bends upwards again towards the Galactic plane,  
traversing the X-ray cluster RXC~J2215+3718 at $(\ell,b,v) \sim (92\degr,-16\degr, 5900\,\kms$) and the cluster ZwCl~2231.2+3732 ($95\degr,-18\degr, 6000\,\kms$) before re-entering the ZoA around $(\ell,b) \sim (90\degr,-10\degr)$. This is also evident in Fig~\ref{fig:pizza}. Most previous studies assumed the PPS stop and dissolve around this location, because no signature of a continuation on the other side of the obscuring ZoA band was seen, not even in the dedicated optical ZoA galaxy searches by \citet{seeberger94}. Our data clearly confirm continuity across the ZoA: on the northern side of the Galactic Plane (around $\ell \sim 80\degr$) the filament seems to continue with a shallower slope, towards an agglomeration of  galaxy groupings centered around $(\ell,b) \sim (50\degr,+15\degr)$ (dashed blue circle in  Fig.~\ref{fig:PPS}) at slightly lower velocities ($\sim 5000\,\kms$).  This continuation across the plane is also visible in Fig.~\ref{fig:pizza} as a wall-like structure within the $b < \pm 10\degr$ redshift slice; here we can trace the wall from $\ell\sim 80\degr$ at 6000\,\kms\ to $\ell\sim 50\degr$ at 4500\,\kms. 

Figure~\ref{fig:PPS} implies the Perseus-Pisces chain to be considerably larger than evidenced in any previous survey of the PPS complex. Extending from the A~569 cluster across the Plane, all the way back to where it re-enters the ZoA, re-surfacing in the northern hemisphere around Cygnus, the feature spans a length of $\Delta \ell \sim 150\degr$ (180\degr\ if we assume its continuation to $\ell \sim 50\degr$ to be real). This apparent contiguous band most likely has its origin in a cut through the surfaces of contiguous galaxy voids in the cosmic web of the nearby Universe, rather than one coherent structure of length of about 400\,$h_{75}^{-1}$\,Mpc. This would also explain the smooth increase and decrease in the average velocity of the PPS wall more readily.
 
\smallskip
\noindent{\bf SuperGalactic Plane and Cassiopeia: } The concentration of detections between $\ell \sim 120\degr - 140\degr$ seems -- upon closer inspection -- mostly due to two distinct features. This is also evident in the redshift shells based on the 2MASX photometric redshift plots \citet{Jarrett2004}, as well as the 2MRS catalog \citep{huchra12} which exclude the inner ZoA. The thin filament around $\ell \sim 140\degr$ is nearer in velocity space ($2500-3500$\,\kms) and forms part of the Supergalactic Plane. The agglomeration around $\ell \sim 130\degr$ is a bit broader, and at slightly higher velocities (around $v \sim 4500$\,\kms), and seems less well defined. According to Fig.~\ref{fig:onsky} it does link to the groupings around ($\ell,b) \sim (140\degr,+20\degr$) , whereas the redshift wedges (Fig.~\ref{fig:pizza}) suggest a wall-like structure that connects the Cas-structure with the PPS ZoA wall discussed above, i.e. extending from $(\ell,v) \sim (120\degr,4500\,\kms)$ to $(165\degr,6500\,\kms)$, despite the low numbers of detections within the longitude range $120\degr-140\degr$.
\smallskip

\noindent{\bf Ophiuchus: } Following a fairly devoid stretch from about $\ell \sim$60\degr\ towards the Galactic Centre -- partly caused by the lack of 2MASS galaxies around the Galactic Centre -- a small peak  appears in the histogram around $\ell=-10\degr$ ($\ell= 350\degr$). This is caused by galaxies in the +5\degr\ to +10\degr\ latitude strip. Figure~\ref{fig:pizza} suggest them to form part of the Ophiuchus supercluster \citep{Wakamatsu1994, Wakamatsu2005} and the outer boundary of the Local Void.

\section{Conclusions and next steps} 

As part of our long-term efforts to map the large-scale
galaxy distribution and associated flow fields behind the Milky Way,  we pursued a dedicated observing campaign with the 100\,m class \nan\ radio telescope to search for the \HI\ emission and measure the redshifts of all bright near-infrared galaxies that so far had none.

Our NRT target list contained 1003 2MASS galaxies brighter than $K^o_s < 11\fm25$ across the ZoA ($|b| < 10\degr$) accessible from \nan\ (Dec\,$ > -40\degr$). The majority of the galaxies without prior redshift measurement are located within the inner ZoA ($|b| < 5\degr$). They were selected from the newly-defined 2MASS ZoA catalogue \citep[2MZOAG,][]{Schroeder2018}. 
  
For the adopted sensitivity limit of rms = 3mJy and the redshift limit of 10\,600\,\kms, the  1800 hrs observing time allocated with the NRT led to the detection of 230 galaxies (of which 10 are marginal). An additional 29 and 6 clear and marginal detections of non-target are presented.

The detection rate of the bright 2MZOAG sample was found to be independent of foreground extinction (Fig.~\ref{fig:AK}), in agreement with what was found in the "blind" HIZoA  survey \citep{hizoa2016}. The 22\% detection rate is consistent with our pilot project \citep{vandriel09}, and quite reasonable considering that no pre-selection on morphological type was made. Moreover, many of the brighter galaxies might have been targeted on earlier occasions.

The galaxies detected in this survey typically are \HI-rich, with \HI-masses ranging in an interval of only \lMHI$ = 8.2 - 10.6$~\Msun. However, only a handful of detections have masses below 9.0\,\Msun. Given the NIR-selection, it is not surprising that their stellar luminosities are fairly high: they hover with a narrow dispersion around a mean of log\,$L_{\rm K} = 10.8$ (see Fig.~\ref{fig:glob}), resulting in a small range in \HI-mass-to-light ratios as well.

The main goal of the \nan\ observations presented here was the unveiling of the large-scale distribution of galaxies in the hitherto unexplored inner ZoA, particularly the northern ZoA in between the Arecibo strips ($70\degr < \ell < 180\degr$) for which hardly any recession velocities existed before (see Figs.~9--11):\\
\indent -- the new \nan\ detections add to our insight into the complex PPS structure. They confirm a continuation of the PPS from the Perseus cluster across the ZoA at $\ell \sim 160\degr$, connecting it to Abell\,569 via the 3C\,129-cluster (see \citealt{ramatsoku16} for WSRT follow-up results of this cluster)\\
\indent -- a further extension shows the PPS to extend further than the Pisces clusters. It can be traced along a number of groups and clusters (see Figs.~9 and 10), folding back into the ZoA in Cygnus ($\ell \sim90\degr$) from where it seems to continue to the northern Galactic hemisphere. As such the PPS spans a length of about 150\degr\ ($\sim 400\,h^{-1}_{75}$\,Mpc) on the sky\\
\indent -- two further filaments crossing the ZoA were identified in Cassiopeia ($\ell \sim 130\degr$) of which one forms part the SGP, with the other being a new feature at a slightly higher distance \\
\indent -- a sprinkling of detections in Ophiuchus (around the Galactic Bulge) appear to form part of the outer boundary of the LG, demonstrating the success of this project.

The results confirm this as a promising avenue to explore the ZoA, because neither the NIR photometry nor the \HI\ radio emission are heavily affected by dust obscuration and stellar confusion. The combination of these data sets will in future allow the derivation of peculiar velocities through the NIR TF-relation \citep[see e.g.][]{Said15TF}, and therewith a better understanding of the local dynamics, the cosmic flow fields and the underlying density field. 

Our parallel projects with the goal of filling the 2MRS- and 2MTF-ZoA are also well underway. Further optical spectroscopic surveys of the 2MZOAG objects  that have bright bulges, but were not detected in \HI, have been launched.  We have nearly completed the \HI-coverage of 2MASX ZoA galaxies for which only optical redshifts or low-quality \HI-profiles existed, and observed the more southern 2MZOAG with Parkes.  And our complimentary approach of obtaining deep follow-up NIR imaging of galaxies detected in "blind" \HI-surveys \citep[e.g][]{hizoa2016} has been completed. This inverse method is particularly successful around the Galactic Bulge, where 2MASX is incomplete.

The above projects can furthermore be regarded as precursors to the forthcoming SKA Pathfinder \HI\ surveys, like WALLABY \citep{koribalski12, duffy12} with ASKAP \citep{johnston08}, APERTIF \citep{oosterloo10} at Westerbork, and MeerKAT \citep{mks2016}, that will all start within the next year and cover major parts of the ZoA to unprecedented depth and resolution. 

\section*{Acknowledgments}

A special word of thanks goes to the late John Huchra with whom we started this project in 2009 given our common interest in unveiling the extragalactic sky covered by the Milky Way. We want to thank Matt Lehnert and Zhon Butcher for the ``derippling'' of affected spectra. RCKK, ACS and MR thank the South African National Research Foundation for support. The \nan\ Radio Telescope is operated as part of the Paris Observatory, in association  with the Centre National de la Recherche Scientifique (CNRS) and partially supported by  the R\'egion Centre in France. This publication makes use of data products from the Two Micron All Sky Survey, which is a joint project  of the University of Massachusetts and the Infrared Processing and Analysis Center, funded by the National Aeronautics and Space Administration and the National Science Foundation. This research also has made use of the HyperLeda database (http://leda.univ-lyon1.fr), the NASA/IPAC Extragalactic Database (NED) which is operated by the Jet Propulsion Laboratory, California Institute of Technology, under contract with the National Aeronautics and Space Administration and the Sloan Digital Sky Survey which is managed by the Astrophysical Research Consortium for the Participating Institutions. 

\footnotesize{
\bibliographystyle{mn2e} 
\bibliography{ZoA_bibfile_281116} 
\label{lastpage}

\end{document}

\appendix

\section{Notes to individual galaxies}
\input{2MASZoA_NRT_paper_app_rev10.tex}

\onecolumn



\newpage
\clearpage 
{\footnotesize
\input{2MASZoA_NRT_paper_table1_HIdets_clear_marginal_rev8}
}

\newpage
\clearpage
{\footnotesize
\input{2MASZoA_NRT_paper_table2_HIdets_nontargets_rev8}
}

\newpage
\clearpage
{\footnotesize
\input{2MASZoA_NRT_paper_table7_HIcomp_rev8}
}

\newpage
\clearpage
{\footnotesize
\input{2MASZoA_NRT_paper_table6_vopt_rev6}
}

\newpage
\clearpage
{\footnotesize
\input{2MASZoA_NRT_paper_table4_nondets_rev9}
}

\newpage
\clearpage
{\footnotesize
\input{2MASZoA_NRT_paper_table5_HIdets_possible_rev5}
}

\newpage
\clearpage
{\footnotesize
\input{2MASZoA_NRT_paper_table3_derivedprops_rev5}
}


\newpage
\clearpage 

\begin{figure*} 
\centering
\includegraphics[width=0.99\textwidth]{spec_H_1.eps}
\caption{\nan\ 21cm \HI\ line spectra of detected galaxies (see Tables
 A1 and A2). Velocity resolution is 18 \kms.}
\label{HIdetspectra1}
\end{figure*}

\newpage
\clearpage
\begin{figure*} 
\centering
\addtocounter{figure}{-1}
\includegraphics[width=0.99\textwidth]{spec_H_2.eps}
\caption{-- {\it continued}.}
\end{figure*}

\newpage
\clearpage
\begin{figure*} 
\centering
\addtocounter{figure}{-1}
\includegraphics[width=0.99\textwidth]{spec_H_3.eps}
\caption{-- {\it continued}.}
\end{figure*}

\newpage
\clearpage
\begin{figure*} 
\centering
\addtocounter{figure}{-1}
\includegraphics[width=0.99\textwidth]{spec_H_4.eps}
\caption{-- {\it continued}.}
\end{figure*}

\newpage
\clearpage
\begin{figure*} 
\centering
\addtocounter{figure}{-1}
\includegraphics[width=0.99\textwidth]{spec_H_5.eps}
\caption{-- {\it continued}.}
\end{figure*}

\newpage
\clearpage
\begin{figure*} 
\centering
\addtocounter{figure}{-1}
\includegraphics[width=0.99\textwidth]{spec_H_6.eps}
\caption{-- {\it continued}.}
\end{figure*}

\newpage
\clearpage
\begin{figure*} 
\addtocounter{figure}{-1}
\noindent\includegraphics[width=0.99\textwidth]{spec_H_7.eps}\\
\caption{-- {\it continued}.}
\end{figure*}

\label{lastpage}

\end{document}

%% file: Table1.tex
\begin{landscape}
\bigskip~\bigskip~\bigskip

\begin{longtable}{llrrrrrrrrrrrrr}
\caption{{\normalsize \HI\ detections -- observational data} \label{tab:detsobsdata}}\\
\hline
\noalign{\smallskip}
2MASX J          & Other name        & $K_{20}$ & $J$-$K$ & $H$-$K$ & $d_{K20}$ & $b/a$  & \vfi  & $\sigma_V$ & \wfi  & \wtw  & \FHI    & $\sigma_F$ & rms  & S/N  \\ 
                 &                   & mag      & mag     &  mag    &$\prime\prime$&     & \KMS\ & \KMS\      & \KMS\ & \KMS\ & \JYKMS\ & \JYKMS\    & mJy  &      \\
(1)              &  (2)              & (3)      & (4)     & (5)     & (6)       & (7)    & (8)   & (9)        & (10)  & (11)  & (12)    & (13)       & (14) & (15) \\
\noalign{\smallskip}
\hline
\noalign{\smallskip}
\endfirsthead
\caption{continued.}\\
\hline
\noalign{\smallskip}
2MASX J          & Other name        & $K_{20}$ & $J$-$K$ & $H$-$K$ & $d_{K20}$ & $b/a$  & \vfi  & $\sigma_V$ & \wfi  & \wtw  & \FHI    & $\sigma_F$ & rms  & S/N  \\ 
                 &                   & mag      & mag     &  mag    &$\prime\prime$&     & \KMS\ & \KMS\      & \KMS\ & \KMS\ & \JYKMS\ & \JYKMS\    & mJy  &      \\
(1)              &  (2)              & (3)      & (4)     & (5)     & (6)       & (7)    & (8)   & (9)        & (10)  & (11)  & (12)    & (13)       & (14) & (15) \\
\noalign{\smallskip}
\hline
\noalign{\smallskip}
\endhead
\noalign{\smallskip}
\hline
\endfoot
00141253+7036448$^{v}$    & PGC 2737274           & 10.92 & 1.37 & 0.39 &  53 & 0.70 & 6958 &  8 & 396 & 430 &  9.08 & 1.09 & 5.66 & 13.4 \\ 
00384223+6017130          & ZOAG 121.35-02.54     & 10.94 & 1.20 & 0.31 &  51 & 0.74 & 4354 &  5 & 222 & 241 &  3.82 & 0.66 & 4.55 &  9.4 \\ 
00475430+6807433$^{a,v}$  & ZOAG 122.61+05.26     & 10.86 & 1.42 & 0.50 &  74 & 0.30 & 3763 &  5 & 341 & 368 & 10.62 & 0.98 & 5.51 & 17.4 \\ 
01191829+6219297          & ZOAG 126.16+00.37     & 10.67 & 1.67 & 0.52 &  62 & 0.22 & 4041 & 18 & 294 & 354 &  4.73 & 0.79 & 4.51 & 10.2 \\ 
01203021+6525055          & IRAS 01170+6509       &  9.96 & 1.43 & 0.45 &  80 & 0.54 & 4148 &  4 & 413 & 440 & 10.64 & 0.87 & 4.48 & 19.5 \\ 
01261932+6046064          & WEIN 013              & 10.15 & 1.43 & 0.48 &  64 & 0.60 & 5958 &  3 & 477 & 490 &  5.12 & 0.55 & 2.69 & 14.5 \\ 
01273787+6308155          & ZOAG 127.01+00.55     & 10.93 & 1.83 & 0.62 &  36 & 0.82 & 6223 & 40 &  95 & 294 &  3.65 & 0.56 & 3.50 & 17.8 \\ 
01485859+6045514          & IRAS 01455+6031       & 11.04 & 1.37 & 0.40 &  46 & 0.40 & 4391 & 11 & 266 & 311 &  5.73 & 0.77 & 4.70 & 12.5 \\ 
01572719+6601408          & IRAS 01536+6546       & 10.57 & 2.23 & 0.72 &  42 & 0.62 & 3905 & 17 & 224 & 292 &  2.62 & 0.48 & 3.03 &  9.6 \\ 
02013241+6824219$^{a,v}$  & IRAS 01575+6809       &  9.59 & 1.37 & 0.41 &  89 & 0.88 & 3771 &  3 &  77 & 102 &  2.20 & 0.25 & 2.67 & 15.7 \\ 
\end{longtable}

%% file: Table2.tex
\bigskip
\bigskip
\bigskip

\begin{longtable}{llrrrrrrrrr}
\caption{{HI detections of non-targets in the telescope beam} \label{tab:nontargetdetsobsdata}}\\
\hline
\noalign{\smallskip}
Target - 2MASX  J    & Detected object        & dist & \vfi  & $\sigma_V$ & \wfi  & \wtw  & \FHI    & $\sigma_F$ & rms  & S/N  \\ 
                     &                    & 0.5 HPBW & \KMS\ & \KMS\      & \KMS\ & \KMS\ & \JYKMS\ & \JYKMS\    & mJy  &      \\
(1)                  &  (2)                   & (3)  & (4)   & (5)        & (6)   & (7)   & (8)     & (9)        & (10) & (11) \\
\noalign{\smallskip}
\hline 
\noalign{\smallskip}
\endfirsthead
\caption{continued.}\\
\hline
\noalign{\smallskip}
Target - 2MASX  J    & Detected object        & dist & \vfi  & $\sigma_V$ & \wfi  & \wtw  & \FHI    & $\sigma_F$ & rms  & S/N  \\ 
                     &                    & 0.5 HPBW & \KMS\ & \KMS\      & \KMS\ & \KMS\ & \JYKMS\ & \JYKMS\    & mJy  &      \\
(1)                  &  (2)                   & (3)  & (4)   & (5)        & (6)   & (7)   & (8)     & (9)        & (10) & (11) \\
\noalign{\smallskip}
\hline
\noalign{\smallskip}
\endhead
\noalign{\smallskip}
\hline
\endfoot
01474890+6305128$^{a,n,v}$ & EZOA J0147+63                 & 2.4 & 4243 & 16 & 189 & 248 &  1.74 & 0.36 & 2.44 &   8.7 \\ 
02085980+7114029$^{a,n}$   & 2MASX J02084091+7102087       & 1.1 & 3304 &  7 & 171 & 209 &  4.47 & 0.61 & 4.51 &  12.6 \\ 
02531475+5528143$^{a,n,c,v}$& Anon J025321.6+553602        & 1.3 & 3832 &  3 &  47 &  86 &  3.33 & 0.28 & 3.19 &  25.4 \\ 
02531969+5529140$^{a,n}$   & Anon J025321.6+553602         & 1.0 & 3824 &  2 &  63 &  97 &  4.92 & 0.29 & 3.14 &  32.9 \\ 
(0253216+553602)$^{s,a,n}$ & Anon J025321.6+553602         & 1.1 & 3826 &  5 &  60 & 113 &  4.26 & 0.41 & 4.13 &  22.2 \\ 
03202205+6645055$^{a,n,v}$ & Anon J031958.7+664959         & 1.3 & 2993 &  3 & 175 & 197 &  4.20 & 0.36 & 2.78 &  19.0 \\ 
03403139+6649043$^{a,n,v}$ & 2MASX J03400768+6642470       & 1.4 & 1574 &  2 & 106 & 165 & 21.60 & 0.36 & 2.98 & 117.3 \\ 
04114143+3841285$^{a,n,v}$ & 2MASX J04113202+3846565       & 1.1 & 5698 & 20 & 102 & 191 &  1.34 & 0.24 & 1.87 &  11.8 \\ 
04580945+3523013$^{a,n}$   & 2MASX J04580771+3533503       & 0.9 & 6277 & 36 & 307 & 427 &  2.55 & 0.40 & 2.06 &  11.8 \\ 
05583605+4031118$^{a,n,v}$ & 2MASX J05583132+4031028       & 0.5 & 7661 & 36 & 207 & 330 &  3.03 & 0.65 & 3.87 &   9.1 \\ 
\end{longtable}
\end{landscape}

%% file: Table3.tex
\begin{table*}
\caption{Comparison with literature HI detections \label{tab:HIdetscomp}}
\begin{tabular}{lrrrrrrrrrl} 
\hline
\noalign{\smallskip}
2MASX J          & \vfi     & \wfi     & \wtw    & \FHI    & \vfi  & \wfi     & \wtw  & \FHI    & Tel  & Ref. \\
                 & \multicolumn{4}{c}{---------- \nan\ ----------} & \multicolumn{4}{c}{---------- literature ----------} & & \\
                 & \KMS\    & \KMS\    & \KMS\   & \JYKMS\ & \KMS\    & \KMS\ & \KMS\ & \JYKMS\ &      &      \\
  (1)            &  (2)     & (3)      & (4)     & (5)     & (6)   & (7)      & (8)   & (9)     & (10) & (11) \\
\noalign{\smallskip}
\hline
00384223+6017130         &  4354 & 222 & 241 &  3.82 & 4355 & 219 & 228 &     * & VLA & H92        \\ 
01191829+6219297         &  4041 & 294 & 354 &  4.73 & 4050 & 306 & 316 &  7.05 & NRT & P03        \\ 
02550583+6624065$^{v}$   &  3487 & 381 & 427 &  6.76 & 3456 & 374 & 390 & 12.31 & JBO & L03        \\ 
(0319587+664959)$^{s,a}$ &  2995 & 175 & 195 & 11.28 & 3005 & 185 & 200 & 16.03 & JBO & L03        \\ 
03290640+6458319$^{v}$   &  2464 & 416 & 446 & 10.53 & 2456 & 411 & 453 & 13.18 & JBO & L03        \\ 
03292042+6601389$^{a,v}$ &  2113 & 349 & 372 &  4.17 & 2111 & 358 & 379 &  4.33 & GBT & M14        \\ 
03480963+4955140$^{a,c,v}$& 9991 & 233 & 453 &  4.19 & 9968 & 351 & 366 &  2.37 & NRT & P03        \\ 
04075531+4549400         &  4470 & 276 & 310 & 11.88 & 4470 & 287 & 320 & 11.48 & NRT & P03        \\ 
\hline
\multicolumn{11}{l}{\bf \footnotesize References:} \\
\multicolumn{11}{p{13cm}}{\scriptsize C99 \citet{chamaraux99}, C09 \citet{courtois09}, D05 \citet{Doyle2005}, H87 \citet{hauschildt87}, 
H88 \citet{haynes88}, 
H92 \citet{henning92}, H00 \citet{henning00}, H10 \citet{henning10}, 
H86 \citet{Huchtmeier86}, K87 \citet{Kerr1987}, K92 \citet{kraan-korteweg92},
K04 \citet{2004BGC}, L90 \citet{lu90}, L03 \citet{lang03}, M90 \citet{martin90}, M04 \citet{Meyer2004}, M14 \citet{masters14}, 
P97 \citet{pantoja97}, P03 \citet{paturel03}, R00 \citet{rosenberg00}, R02 \citet{ryan-weber02}, 
S94 \citet{seeberger94}, S05 \citet{springob05}, S16 \citet{hizoa2016}, W06 \citet{Wong2006}. } 

\end{tabular}
\end{table*}
%

%% file: Table4.tex
\begin{table}
\caption{Optical velocities in the literature \label{tab:optvel}} 
\begin{tabular}{lcrrl}
\hline
2MASX J               & $\Delta v$-flag & \vopt & $\sigma_v$ & Ref.  \\
                      &                 & \KMS\ & \KMS\      &       \\
(1)                   & (2)             & (3)   & (4)        & (5)   \\
\noalign{\smallskip}
\hline
00141253+7036448        &   &   6973\phantom{$^x$} & 43 & H12 \\
00253292+6821442        &   &   3598\phantom{$^x$} &    & M08 \\
                        &   &   3729\phantom{$^x$} & 54 & H12 \\
00475430+6807433$^a$    & * &   1273\phantom{$^x$} & 65 & H12 \\
01311294+6735115        &   &  10345\phantom{$^x$} & 46 & H12 \\
01474890+6305128$^{a}$  &   &   4211\phantom{$^x$} & 58 & H12 \\
01582742+6744421$^{v+}$ &   &  31270\phantom{$^x$} & 44 & H12 \\
02013241+6824219$^a$    & * &   4554\phantom{$^x$} & 55 & F95 \\
02021798+6721240        &   &   3762\phantom{$^x$} & 25 & H12 \\
02034762+6843532$^r$    &   &   9321\phantom{$^x$} & 18 & H12 \\
\hline                                                            
\multicolumn{5}{l}{\bf \footnotesize References:} \\              
\multicolumn{5}{p{8.0cm}}{\scriptsize B09 \citet{beckmann09}, C08 \citet{crook08}, D90 \citet{djorgovski90}, D91 \citet{dressler91}, F95 \citet{fisher95}, G08 \citet{goncalves08}, H12 \citet{huchra12}, H91 \citet{hewitt91}, H95 \citet{huchra95}, J09 \citet{jones09}, K11 \citet{koss11}, L99 \citet{lawrence99}, M96 \citet{marzke96}, M98 \citet{motch98}, N97 \citet{nakanishi97}, P09 \citet{parisi09}, R11 \citet{ricci11}, S85 \citet{spinrad85},  S92 \citet{strauss92}, S98 \citet{seeberger98}, S00 \citet{saunders00}, T94 \citet{takata94}, V96 \citet{visvanathan96}, Y93 \citet{yamada93},  Y94 \citet{yamada94}. } \\     
\end{tabular}
\end{table}                                               

%% file: Table5.tex
\begin{table*}
\begin{center} 
\caption{{HI non-detections -- observational data} \label{tab:nondetsobsprops}}
\begin{tabular}{llrrrrrrrr}
\hline
\noalign{\smallskip}
2MASX J          & Other name & $l$ & $b$ & $K_{20}$ & $A_K$ & \koc         & \jkoc       & \hkoc       & rms  \\ 
                 &            & deg & deg & mag      & mag   & mag          & mag         & mag         & mJy  \\
(1)              & (2)        & (3) & (4) & (5)      & (6)   & (7)          & (8)         & (9)         & (10) \\
\noalign{\smallskip}
\hline
00161976+7025219          & *                  & 119.97 &  7.75 & 11.33 &  0.32 & 10.95 &   0.54 &   0.17 & 3.54 \\ 
00223972+6139447          & *                  & 119.52 & -1.02 & 11.49 &  0.54 & 10.77 &   0.89 &   0.23 & 2.84 \\ 
00253292+6821442$^{v}$    & ZOAG G120.54+05.61 & 120.54 &  5.61 & 10.05 &  0.32 &  9.67 &   0.92 &   0.25 & 3.31 \\ 
00281959+6447011          & *                  & 120.47 &  2.02 & 11.84 &  1.01 & 10.52 &   0.51 &   0.16 & 3.52 \\ 
00303425+6444585$^d$      & *                  & 120.71 &  1.97 & 11.82 &  0.54 & 11.14 &   1.05 &   0.24 & 2.31 \\ 
00314802+6227440          & *                  & 120.66 & -0.32 & 11.88 &  0.61 & 11.08 &      * &      * & 3.07 \\ 
00321813+6029565          & ZOAG G120.58-02.29 & 120.58 & -2.29 & 11.38 &  0.25 & 11.09 &   0.92 &   0.23 & 2.53 \\ 
00343086+6257451          & ZOAG G121.01+00.15 & 121.01 &  0.15 & 11.23 &  0.45 & 10.67 &   0.65 &   0.12 & 2.11 \\ 
00343656+6310260          & *                  & 121.04 &  0.36 & 11.53 &  0.47 & 10.95 &   0.91 &   0.19 & 2.93 \\ 
00360888+6319171          & *                  & 121.22 &  0.50 & 11.37 &  0.42 & 10.86 &   1.36 &   0.49 & 2.20 \\ 
\hline
\end{tabular}
\end{center}
\end{table*}

%% file: Table6.tex
\begin{landscape}
\bigskip~\bigskip

\begin{longtable}{llrrrrrrrrrrrrr}
\caption{{Possible HI detections -- observational data} \label{tab:possdetsobsdata}}\\
\hline
\noalign{\smallskip}
2MASX J          & Other name        & $K_{20}$ & $J$-$K$ & $H$-$K$ & $d_{K20}$ & $b/a$  & \vfi  & $\sigma_V$ & \wfi  & \wtw  & \FHI    & $\sigma_F$ & rms  & S/N  \\ 
                 &                   & mag      & mag     &  mag    & $''$      &        & \KMS\ & \KMS\      & \KMS\ & \KMS\ & \JYKMS\ & \JYKMS\    & mJy  &      \\
(1)              &  (2)              & (3)      & (4)     & (5)     & (6)       & (7)    & (8)   & (9)        & (10)  & (11)  & (12)    & (13)       & (14) & (15) \\
\noalign{\smallskip}
\hline
\noalign{\smallskip}
\endfirsthead
\caption{continued.}\\
\hline
\noalign{\smallskip}
2MASX J          & Other name        & $K_{20}$ & $J$-$K$ & $H$-$K$ & $d_{K20}$ & $b/a$  & \vfi  & $\sigma_V$ & \wfi  & \wtw  & \FHI    & $\sigma_F$ & rms  & S/N  \\ 
                 &                   & mag      & mag     &  mag    & $''$      &        & \KMS\ & \KMS\      & \KMS\ & \KMS\ & \JYKMS\ & \JYKMS\    & mJy  &      \\
(1)              &  (2)              & (3)      & (4)     & (5)     & (6)       & (7)    & (8)   & (9)        & (10)  & (11)  & (12)    & (13)       & (14) & (15) \\
\noalign{\smallskip}
\hline
\noalign{\smallskip}
\endhead
\noalign{\smallskip}
\hline
\endfoot
02590152+5318199            & ZOAG G141.41-04.92 & 11.49 & 1.35 & 0.38 &  38 & 0.32 & 7146 &  8 & 229 & 250 & 0.79 & 0.30 & 2.05 &  4.3 \\ 
03104409+6106477$^{a}$      & ZOAG G138.96+02.65 &  8.08 & 1.51 & 0.48 & 133 & 0.54 & 2505 & 38 & 162 & 267 & 1.47 & 0.41 & 2.73 &  7.1 \\ 
03192996+5755423            & *                  & 11.61 & 1.39 & 0.48 &  24 & 0.78 & 1327 & 15 & 254 & 291 & 1.55 & 0.41 & 2.58 &  6.3 \\ 
04110830+3837269B           & *                  & 10.90 & 1.45 & 0.45 &  46 & 0.34 & 6158 & 13 & 234 & 273 & 1.62 & 0.31 & 2.03 &  8.7 \\ 
04264449+3810182            & 2MFGC 03603        & 10.87 & 2.07 & 0.65 &  61 & 0.30 & 6546 & 32 & 129 & 216 & 0.82 & 0.24 & 1.77 &  6.8 \\ 
04292626+4855120            & ZOAG G155.28+00.24 & 10.87 & 1.62 & 0.53 &  45 & 0.42 & 7231 & 30 &  94 & 154 & 0.55 & 0.25 & 2.18 &  4.3 \\ 
04413675+4203562            & *                  & 11.13 & 1.44 & 0.38 &  38 & 0.48 & 3975 & 22 &  99 & 147 & 0.36 & 0.14 & 1.27 &  4.7 \\ 
05141860+4622066            & ZOAG G162.04+04.44 & 10.73 & 1.25 & 0.31 &  61 & 0.46 & 6388 &  8 & 388 & 405 & 1.21 & 0.36 & 1.90 &  5.4 \\ 
06162583+1654326            & *                  & 11.59 & 1.60 & 0.53 &  38 & 0.60 & 5131 & 54 &  55 & 239 & 1.18 & 0.33 & 2.30 & 11.5 \\ 
18202335-0117447            & 2MFGC 14346        & 11.47 & 1.69 & 0.51 &  53 & 0.20 & 6875 & 11 & 455 & 484 & 1.63 & 0.44 & 2.14 &  6.0 \\ 
20135618+4443093$^{v}$      & 2MIG 2761          & 11.29 & 1.52 & 0.50 &  45 & 0.42 & 7078 &  9 & 360 & 383 & 1.69 & 0.40 & 2.22 &  6.7 \\ 
21071353+4456529$^{c?}$     & *                  & 10.34 & 1.58 & 0.45 &  50 & 0.90 & 4793 & 28 & 460 & 535 & 3.13 & 0.59 & 2.75 &  8.8 
\end{longtable}

%% file: Table7.tex
\bigskip~\bigskip

\begin{longtable}{lrrrrrrrrrrrrrr}
  \caption{{Sample of HI detections -- derived properties} \label{tab:detsderprops}}\\
\hline
\noalign{\smallskip}
2MASX J & $l$ & $b$ & $A_K$ & $D$ & \koc       & \jkoc     & \hkoc     & \rkc & \lLKc   & \lMHI  & $\log({M_{\rmn{HI}}\over L_{K_{c}}})$ & \vrot & \lMdyn & \lMbardyn \\
        & deg & deg & mag   & Mpc & mag        & mag       & mag       & kpc  & \LsunK\ & \Msun\ & \Msun/\LsunK\                & \KMS\ & \Msun\ &           \\
(1)     & (2) & (3) & (4)   & (5) & (6)        & (7)       & (8)       & (9)  & (10)    & (11)   & (12)                                  & (13)  & (14)   & (15)      \\
\noalign{\smallskip}
\hline
00141253+7036448$^{v}$     & 119.82 &  7.97 & 0.31 &  92.6 & 10.54 &   0.93 &   0.24 &  9.9 & 11.04 & 10.26 & -0.78 & 277 & 11.25 & -0.19 \\ 
00384223+6017130           & 121.35 & -2.55 & 0.22 &  55.0 & 10.67 &   0.89 &   0.21 &  5.7 & 10.54 &  9.43 & -1.10 & 165 & 10.56 & -0.06 \\ 
00475430+6807433$^{a,v}$   & 122.60 &  5.26 & 0.35 &  50.0 & 10.37 &   0.92 &   0.33 &  7.7 & 10.57 &  9.80 & -0.78 & 179 & 10.76 & -0.17 \\ 
01191829+6219297           & 126.16 & -0.37 & 0.38 &  55.7 & 10.22 &   1.14 &   0.34 &  6.8 & 10.73 &  9.54 & -1.19 & 151 & 10.55 &  0.12 \\ 
01203021+6525055           & 125.95 &  2.72 & 0.40 &  54.3 &  9.47 &   0.88 &   0.26 &  9.2 & 11.01 &  9.87 & -1.14 & 245 & 11.11 & -0.15 \\ 
01261932+6046064           & 127.18 & -1.82 & 0.21 &  78.5 &  9.92 &   1.14 &   0.38 &  9.6 & 11.15 &  9.87 & -1.27 & 298 & 11.30 & -0.21 \\ 
01273787+6308155           & 127.01 &  0.55 & 0.53 &  86.3 & 10.28 &   1.09 &   0.37 &  6.4 & 11.08 &  9.81 & -1.28 &  83 & 10.01 &  1.01 \\ 
01485859+6045514           & 129.90 & -1.33 & 0.35 &  57.7 & 10.61 &   0.88 &   0.24 &  5.4 & 10.60 &  9.65 & -0.95 & 145 & 10.42 &  0.16 \\ 
01572719+6601408           & 129.58 &  4.01 & 0.41 &  53.9 & 10.08 &   1.66 &   0.52 &  4.5 & 10.75 &  9.25 & -1.50 & 143 & 10.33 &  0.35 \\ 
02013241+6824219$^{a,v}$   & 129.34 &  6.41 & 0.35 &  48.3 &  9.19 &   0.89 &   0.25 &  8.9 & 11.02 &  9.08 & -1.93 &  81 & 10.13 &  0.79 \\ 
\hline
\end{longtable}
\end{landscape}

%% file: 2M_ZOA_MN.bbl
\begin{thebibliography}{132}
\expandafter\ifx\csname natexlab\endcsname\relax\def\natexlab#1{#1}\fi

\bibitem[{{Beckmann} {et~al}\mbox{.}(2009){Beckmann}, {Soldi}, {Ricci},
  {Alfonso-Garz{\'o}n}, {Courvoisier}, {Domingo}, {Gehrels}, {Lubi{\'n}ski},
  {Mas-Hesse}, \& {Zdziarski}}]{beckmann09}
{Beckmann} V. {et~al.}, 2009, A\&A, 505, 417

\bibitem[{{Butcher} {et~al}\mbox{.}(2016){Butcher}, {Schneider}, {van Driel},
  {Lehnert}, \& {Minchin}}]{butcher16}
{Butcher} Z., {Schneider} S., {van Driel} W., {Lehnert} M.~D., {Minchin} R.,
  2016, \aap, 596, A60

\bibitem[{{Carrick} {et~al}\mbox{.}(2015){Carrick}, {Turnbull}, {Lavaux}, \&
  {Hudson}}]{carrick2015}
{Carrick} J., {Turnbull} S.~J., {Lavaux} G., {Hudson} M.~J., 2015, \mnras, 450,
  317

\bibitem[{{Chamaraux} {et~al}\mbox{.}(1990){Chamaraux}, {Cayatte}, {Balkowski},
  \& {Fontanelli}}]{Chamaraux1990}
{Chamaraux} P., {Cayatte} V., {Balkowski} C., {Fontanelli} P., 1990, A\&A, 229,
  340

\bibitem[{{Chamaraux} {et~al}\mbox{.}(1999){Chamaraux}, {Masnou}, {Kaz{\'e}s},
  {Sait{\= o}}, {Takata}, \& {Yamada}}]{chamaraux99}
{Chamaraux} P., {Masnou} J.-L., {Kaz{\'e}s} I., {Sait{\= o}} M., {Takata} T.,
  {Yamada} T., 1999, MNRAS, 307, 236

\bibitem[{{Colina} \& {Bohlin}(1997)}]{colina97}
{Colina} L., {Bohlin} R., 1997, \aj, 113, 1138

\bibitem[{{Courtois} {et~al}\mbox{.}(2009){Courtois}, {Tully}, {Fisher},
  {Bonhomme}, {Zavodny}, \& {Barnes}}]{courtois09}
{Courtois} H.~M., {Tully} R.~B., {Fisher} J.~R., {Bonhomme} N., {Zavodny} M.,
  {Barnes} A., 2009, \aj, 138, 1938

\bibitem[{{Crook} {et~al}\mbox{.}(2008){Crook}, {Huchra}, {Martimbeau},
  {Masters}, {Jarrett}, \& {Macri}}]{crook08}
{Crook} A.~C., {Huchra} J.~P., {Martimbeau} N., {Masters} K.~L., {Jarrett} T.,
  {Macri} L.~M., 2008, \apj, 685, 1320

\bibitem[{{Djorgovski} {et~al}\mbox{.}(1990){Djorgovski}, {Thompson}, {de
  Carvalho}, \& {Mould}}]{djorgovski90}
{Djorgovski} S., {Thompson} D.~J., {de Carvalho} R.~R., {Mould} J.~R., 1990,
  \aj, 100, 599

\bibitem[{{Donley} {et~al}\mbox{.}(2005){Donley} {et~al.}}]{Donley2005}
{Donley} J.~L., {et~al.}, 2005, AJ, 129, 220

\bibitem[{{Doyle} {et~al}\mbox{.}(2005){Doyle}, {Drinkwater}, {Rohde},
  {Pimbblet}, {Read}, {Meyer}, {Zwaan}, {Ryan-Weber}, {et~al.}}]{Doyle2005}
{Doyle} M.~T. {et~al.}, 2005, MNRAS, 361, 34

\bibitem[{{Dressler}(1991)}]{dressler91}
{Dressler} A., 1991, \apjs, 75, 241

\bibitem[{{Dressler} {et~al}\mbox{.}(1987){Dressler}, {Lynden-Bell},
  {Burstein}, {Davies}, {Faber}, {Terlevich}, \& {Wegner}}]{Dressler1987}
{Dressler} A., {Lynden-Bell} D., {Burstein} D., {Davies} R.~L., {Faber} S.~M.,
  {Terlevich} R., {Wegner} G., 1987, ApJ, 313, 42

\bibitem[{{Duffy} {et~al}\mbox{.}(2012){Duffy}, {Meyer}, {Staveley-Smith},
  {Bernyk}, {Croton}, {Koribalski}, {Gerstmann}, \& {Westerlund}}]{duffy12}
{Duffy} A.~R., {Meyer} M.~J., {Staveley-Smith} L., {Bernyk} M., {Croton} D.~J.,
  {Koribalski} B.~S., {Gerstmann} D., {Westerlund} S., 2012, \mnras, 426, 3385

\bibitem[{{Ebeling} {et~al}\mbox{.}(2000){Ebeling}, {Edge}, {Allen},
  {Crawford}, {Fabian}, \& {Huchra}}]{ebeling00}
{Ebeling} H., {Edge} A.~C., {Allen} S.~W., {Crawford} C.~S., {Fabian} A.~C.,
  {Huchra} J.~P., 2000, \mnras, 318, 333

\bibitem[{{Erdo\u{g}du} {et~al}\mbox{.}(2006){Erdo\u{g}du}
  {et~al.}}]{erdogdu2006}
{Erdo\u{g}du} P., {et~al.}, 2006, MNRAS, 368, 1515

\bibitem[{{Fisher} {et~al}\mbox{.}(1995){Fisher}, {Huchra}, {Strauss}, {Davis},
  {Yahil}, \& {Schlegel}}]{fisher95}
{Fisher} K.~B., {Huchra} J.~P., {Strauss} M.~A., {Davis} M., {Yahil} A.,
  {Schlegel} D., 1995, ApJS, 100, 69

\bibitem[{{Fitzpatrick}(1999)}]{Fitzpatrick99}
{Fitzpatrick} E.~L., 1999, \pasp, 111, 63

\bibitem[{{Fixsen} {et~al}\mbox{.}(1996){Fixsen}, {Cheng}, {Gales}, {Mather},
  {Shafer}, \& {Wright}}]{Fixsen1996}
{Fixsen} D.~J., {Cheng} E.~S., {Gales} J.~M., {Mather} J.~C., {Shafer} R.~A.,
  {Wright} E.~L., 1996, ApJ, 473, 576

\bibitem[{{Focardi}, {Marano} \& {Vettolani}(1984){Focardi}, {Marano}, \&
  {Vettolani}}]{focardi84}
{Focardi} P., {Marano} B., {Vettolani} G., 1984, \aap, 136, 178

\bibitem[{{Focardi}, {Marano} \& {Vettolani}(1986){Focardi}, {Marano}, \&
  {Vettolani}}]{focardi86}
{Focardi} P., {Marano} B., {Vettolani} G., 1986, A\&A, 161, 217

\bibitem[{{Giovanelli} \& {Haynes}(1982)}]{giovanelli1982}
{Giovanelli} R., {Haynes} M.~P., 1982, \aj, 87, 1355

\bibitem[{{Goncalves} {et~al}\mbox{.}(2008){Goncalves}, {Martin}, {Halpern},
  {Eracleous}, \& {Pavlov}}]{goncalves08}
{Goncalves} T.~S., {Martin} D.~C., {Halpern} J.~P., {Eracleous} M., {Pavlov}
  G.~G., 2008, The Astronomer's Telegram, 1623

\bibitem[{{Gregory}, {Thompson} \& {Tifft}(1981){Gregory}, {Thompson}, \&
  {Tifft}}]{gregory1981}
{Gregory} S.~A., {Thompson} L.~A., {Tifft} W.~G., 1981, \apj, 243, 411

\bibitem[{{Hauschildt}(1987)}]{hauschildt87}
{Hauschildt} M., 1987, A\&A, 184, 43

\bibitem[{{Haynes} {et~al}\mbox{.}(2011){Haynes}, {Giovanelli}, {Martin},
  {Hess}, {Saintonge}, {Adams}, {Hallenbeck}, {Hoffman}, {Huang}, {Kent},
  {Koopmann}, {Papastergis}, {Stierwalt}, {Balonek}, {Craig}, {Higdon},
  {Kornreich}, {Miller}, {O'Donoghue}, {Olowin}, {Rosenberg}, {Spekkens},
  {Troischt}, \& {Wilcots}}]{haynes11}
{Haynes} M.~P. {et~al.}, 2011, \aj, 142, 170

\bibitem[{{Haynes} {et~al}\mbox{.}(1988){Haynes}, {Magri}, {Giovanelli}, \&
  {Starosta}}]{haynes88}
{Haynes} M.~P., {Magri} C., {Giovanelli} R., {Starosta} B.~M., 1988, AJ, 95,
  607

\bibitem[{{Henning}(1992)}]{henning92}
{Henning} P.~A., 1992, ApJS, 78, 365

\bibitem[{{Henning}(1997)}]{Henning1997}
{Henning} P.~A., 1997, Publications of the Astronomical Society of Australia,
  14, 21

\bibitem[{{Henning} {et~al}\mbox{.}(2008){Henning}, {Springob}, {Day},
  {Minchin}, {Momjian}, {Catinella}, {Muller}, {Koribalski}, {Masters},
  {Pantoja}, {Putman}, {Rosenberg}, {Schneider}, \&
  {Staveley-Smith}}]{henning08}
{Henning} P.~A. {et~al.}, 2008, in American Institute of Physics Conference
  Series, Vol. 1035, The Evolution of Galaxies Through the Neutral Hydrogen
  Window, {Minchin} R., {Momjian} E., eds., pp. 246--248

\bibitem[{{Henning} {et~al}\mbox{.}(2010){Henning}, {Springob}, {Minchin},
  {Momjian}, {Catinella}, {McIntyre}, {Day}, {Muller}, {Koribalski},
  {Rosenberg}, {Schneider}, {Staveley-Smith}, \& {van Driel}}]{henning10}
{Henning} P.~A. {et~al.}, 2010, AJ, 139, 2130

\bibitem[{{Henning} {et~al}\mbox{.}(2000){Henning}, {Staveley-Smith}, {Ekers},
  {Green}, {Haynes}, {Juraszek}, {Kesteven}, {Koribalski}, {Kraan-Korteweg},
  {Price}, {Sadler}, \& {Schr{\"o}der}}]{henning00}
{Henning} P.~A. {et~al.}, 2000, AJ, 119, 2686

\bibitem[{{Hewitt} \& {Burbidge}(1991)}]{hewitt91}
{Hewitt} A., {Burbidge} G., 1991, \apjs, 75, 297

\bibitem[{{Hoffman}, {Courtois} \& {Tully}(2015){Hoffman}, {Courtois}, \&
  {Tully}}]{hoffman2015}
{Hoffman} Y., {Courtois} H.~M., {Tully} R.~B., 2015, \mnras, 449, 4494

\bibitem[{{Hong} {et~al}\mbox{.}(2014){Hong}, {Springob}, {Staveley-Smith},
  {Scrimgeour}, {Masters}, {Macri}, {Koribalski}, {Jones}, \&
  {Jarrett}}]{Hong2014}
{Hong} T. {et~al.}, 2014, \mnras, 445, 402

\bibitem[{{Hong} {et~al}\mbox{.}(2013){Hong}, {Staveley-Smith}, {Masters},
  {Springob}, {Macri}, {Koribalski}, {Jones}, {Jarrett}, \& {Crook}}]{Hong2013}
{Hong} T. {et~al.}, 2013, \mnras, 432, 1178

\bibitem[{{Howlett} {et~al}\mbox{.}(2017){Howlett}, {Staveley-Smith}, {Elahi},
  {Hong}, {Jarrett}, {Jones}, {Koribalski}, {Macri}, {Masters}, \&
  {Springob}}]{Howlett2017}
{Howlett} C. {et~al.}, 2017, \mnras, 471, 3135

\bibitem[{{Huchra} {et~al}\mbox{.}(2005){Huchra}, {Jarrett}, {Skrutskie},
  {Cutri}, {Schneider}, {Macri}, {Steining}, {Mader}, {Martimbeau}, \&
  {George}}]{huchra05}
{Huchra} J. {et~al.}, 2005, in Astronomical Society of the Pacific Conference
  Series, Vol. 329, Nearby Large-Scale Structures and the Zone of Avoidance,
  {Fairall} A.~P., {Woudt} P.~A., eds., p. 135

\bibitem[{{Huchra}, {Geller} \& {Corwin}(1995){Huchra}, {Geller}, \&
  {Corwin}}]{huchra95}
{Huchra} J.~P., {Geller} M.~J., {Corwin}, Jr. H.~G., 1995, ApJS, 99, 391

\bibitem[{{Huchra} {et~al}\mbox{.}(2012){Huchra}, {Macri}, {Masters},
  {Jarrett}, {Berlind}, \& {Calkins}}]{huchra12}
{Huchra} J.~P., {Macri} L.~M., {Masters} K.~L., {Jarrett} T.~H., {Berlind} P.,
  {Calkins} M., 2012, ApJS, 199, 26

\bibitem[{{Huchtmeier} \& {Richter}(1986)}]{Huchtmeier86}
{Huchtmeier} W.~K., {Richter} O.~G., 1986, \aaps, 63, 323

\bibitem[{{Hudson} {et~al}\mbox{.}(2004){Hudson}, {Smith}, {Lucey}, \&
  {Branchini}}]{hudson2004}
{Hudson} M.~J., {Smith} R.~J., {Lucey} J.~R., {Branchini} E., 2004, \mnras,
  352, 61

\bibitem[{{Jarrett}(2004)}]{Jarrett2004}
{Jarrett} T.~H., 2004, Publications of the Astronomical Society of Australia,
  21, 396

\bibitem[{{Jarrett} {et~al}\mbox{.}(2000){Jarrett}, {Chester}, {Cutri},
  {Schneider}, {Skrutskie}, \& {Huchra}}]{Jarrett+2000a}
{Jarrett} T.~H., {Chester} T., {Cutri} R., {Schneider} S., {Skrutskie} M.,
  {Huchra} J.~P., 2000, AJ, 119, 2498

\bibitem[{{Jarrett} {et~al}\mbox{.}(2003){Jarrett}, {Chester}, {Cutri},
  {Schneider}, \& {Huchra}}]{jarrett03}
{Jarrett} T.~H., {Chester} T., {Cutri} R., {Schneider} S.~E., {Huchra} J.~P.,
  2003, AJ, 125, 525

\bibitem[{{Johnston} {et~al}\mbox{.}(2008){Johnston}, {Taylor}, {Bailes},
  {Bartel}, {Baugh}, {Bietenholz}, {Blake}, {Braun}, {Brown}, {Chatterjee},
  {Darling}, {Deller}, {Dodson}, {Edwards}, {Ekers}, {Ellingsen}, {Feain},
  {Gaensler}, {Haverkorn}, {Hobbs}, {Hopkins}, {Jackson}, {James}, {Joncas},
  {Kaspi}, {Kilborn}, {Koribalski}, {Kothes}, {Landecker}, {Lenc}, {Lovell},
  {Macquart}, {Manchester}, {Matthews}, {McClure-Griffiths}, {Norris}, {Pen},
  {Phillips}, {Power}, {Protheroe}, {Sadler}, {Schmidt}, {Stairs},
  {Staveley-Smith}, {Stil}, {Tingay}, {Tzioumis}, {Walker}, {Wall}, \&
  {Wolleben}}]{johnston08}
{Johnston} S. {et~al.}, 2008, Experimental Astronomy, 22, 151

\bibitem[{{Jones} {et~al}\mbox{.}(2009{\natexlab{a}}){Jones}, {Read},
  {Saunders}, {Colless}, \& {Jarrett}}]{jones09}
{Jones} D.~H., {Read} M.~A., {Saunders} W., {Colless} M., {Jarrett} T.,
  2009{\natexlab{a}}, MNRAS, 399, 683

\bibitem[{{Jones} {et~al}\mbox{.}(2009{\natexlab{b}}){Jones}, {Read},
  {Saunders}, {Colless}, {Jarrett}, {Parker}, {Fairall}, {Mauch}, {Sadler},
  {Watson}, {Burton}, {Campbell}, {Cass}, {Croom}, {Dawe}, {Fiegert},
  {Frankcombe}, {Hartley}, {Huchra}, {James}, {Kirby}, {Lahav}, {Lucey},
  {Mamon}, {Moore}, {Peterson}, {Prior}, {Proust}, {Russell}, {Safouris},
  {Wakamatsu}, {Westra}, \& {Williams}}]{jones2009}
{Jones} D.~H. {et~al.}, 2009{\natexlab{b}}, \mnras, 399, 683

\bibitem[{{Kerp} {et~al}\mbox{.}(2011){Kerp}, {Winkel}, {Ben Bekhti},
  {Fl{\"o}er}, \& {Kalberla}}]{kerp11}
{Kerp} J., {Winkel} B., {Ben Bekhti} N., {Fl{\"o}er} L., {Kalberla} P.~M.~W.,
  2011, Astronomische Nachrichten, 332, 637

\bibitem[{{Kerr} \& {Henning}(1987)}]{Kerr1987}
{Kerr} F.~J., {Henning} P.~A., 1987, ApJl, 320, L99

\bibitem[{{Kocevski} \& {Ebeling}(2006)}]{kocevski2006}
{Kocevski} D.~D., {Ebeling} H., 2006, \apj, 645, 1043

\bibitem[{{Koribalski}(2012)}]{koribalski12}
{Koribalski} B.~S., 2012, PASA, 29, 359

\bibitem[{{Koribalski} {et~al}\mbox{.}(2004){Koribalski}, {Staveley-Smith},
  {Kilborn}, {Ryder}, {Kraan-Korteweg}, {Ryan-Weber}, {Ekers}, {Jerjen},
  {Henning}, {Putman}, {Zwaan}, {de Blok}, {Calabretta}, {Disney}, {Minchin},
  {Bhathal}, {Boyce}, {Drinkwater}, {Freeman}, {Gibson}, {Green}, {Haynes},
  {Juraszek}, {Kesteven}, {Knezek}, {Mader}, {Marquarding}, {Meyer}, {Mould},
  {Oosterloo}, {O'Brien}, {Price}, {Sadler}, {Schr{\"o}der}, {Stewart},
  {Stootman}, {Waugh}, {Warren}, {Webster}, \& {Wright}}]{2004BGC}
{Koribalski} B.~S. {et~al.}, 2004, \aj, 128, 16

\bibitem[{{Koss} {et~al}\mbox{.}(2011){Koss}, {Mushotzky}, {Veilleux},
  {Winter}, {Baumgartner}, {Tueller}, {Gehrels}, \& {Valencic}}]{koss11}
{Koss} M., {Mushotzky} R., {Veilleux} S., {Winter} L.~M., {Baumgartner} W.,
  {Tueller} J., {Gehrels} N., {Valencic} L., 2011, \apj, 739, 57

\bibitem[{{Kraan-Korteweg}(2005)}]{KK2005}
{Kraan-Korteweg} R.~C., 2005, in Reviews in Modern Astronomy, Vol.~18, Reviews
  in Modern Astronomy, {R{\"o}ser} S., ed., pp. 48--75

\bibitem[{{Kraan-Korteweg} {et~al}\mbox{.}(2017){Kraan-Korteweg}, {Cluver},
  {Bilicki}, {Jarrett}, {Colless}, {Elagali}, {B{\"o}hringer}, \&
  {Chon}}]{KK2017}
{Kraan-Korteweg} R.~C., {Cluver} M.~E., {Bilicki} M., {Jarrett} T.~H.,
  {Colless} M., {Elagali} A., {B{\"o}hringer} H., {Chon} G., 2017, \mnras, 466,
  L29

\bibitem[{{Kraan-Korteweg} {et~al}\mbox{.}(2016){Kraan-Korteweg}, {Elson},
  {Blyth}, {Carignan}, {Frank}, {Jarrett}, {Cluver}, {Serra}, \&
  {Jozsa}}]{mks2016}
{Kraan-Korteweg} R.~C. {et~al.}, 2016, On the Pathway to the SKA. (Proc. of
  Science: MeerKAT2016), 21

\bibitem[{{Kraan-Korteweg}, {Henning} \& {Schr{\"o}der}(2002){Kraan-Korteweg},
  {Henning}, \& {Schr{\"o}der}}]{2002A&A...391..887K}
{Kraan-Korteweg} R.~C., {Henning} P.~A., {Schr{\"o}der} A.~C., 2002, \aap, 391,
  887

\bibitem[{{Kraan-Korteweg} \& {Huchtmeier}(1992)}]{kraan-korteweg92}
{Kraan-Korteweg} R.~C., {Huchtmeier} W.~K., 1992, A\&A, 266, 150

\bibitem[{{Kraan-Korteweg} {et~al}\mbox{.}(2015){Kraan-Korteweg}, {Jarrett},
  {Elagali}, {Cluver}, {Bilicki}, \& {Colless}}]{rkk_salt_15}
{Kraan-Korteweg} R.~C., {Jarrett} T.~H., {Elagali} A., {Cluver} M.~E.,
  {Bilicki} M., {Colless} M.~M., 2015, SALT Science Conference 2015 (roc. of
  Science: SSC2015), 40

\bibitem[{{Kraan-Korteweg} \& {Lahav}(2000)}]{KKLahav2000}
{Kraan-Korteweg} R.~C., {Lahav} O., 2000, A\&Ar, 10, 211

\bibitem[{{Kraan-Korteweg} {et~al}\mbox{.}(2008){Kraan-Korteweg}, {Shafi},
  {Koribalski}, {Staveley-Smith}, {Buckland}, {Henning}, \&
  {Fairall}}]{KKLV2008}
{Kraan-Korteweg} R.~C., {Shafi} N., {Koribalski} B.~S., {Staveley-Smith} L.,
  {Buckland} P., {Henning} P.~A., {Fairall} A.~P., 2008, Astrophysics and Space
  Science Proceedings, 5, 13

\bibitem[{{Kraan-Korteweg} {et~al}\mbox{.}(1996){Kraan-Korteweg}, {Woudt},
  {Cayatte}, {Fairall}, {Balkowski}, \& {Henning}}]{KK1996}
{Kraan-Korteweg} R.~C., {Woudt} P.~A., {Cayatte} V., {Fairall} A.~P.,
  {Balkowski} C., {Henning} P.~A., 1996, \nat, 379, 519

\bibitem[{{Lang} {et~al}\mbox{.}(2003){Lang}, {Boyce}, {Kilborn}, {Minchin},
  {Disney}, {Jordan}, {Grossi}, {Garcia}, {Freeman}, {Phillipps}, \&
  {Wright}}]{lang03}
{Lang} R.~H. {et~al.}, 2003, VizieR Online Data Catalog, 734, 20738

\bibitem[{{Lawrence} {et~al}\mbox{.}(1999){Lawrence}, {Rowan-Robinson},
  {Ellis}, {Frenk}, {Efstathiou}, {Kaiser}, {Saunders}, {Parry}, {Xiaoyang}, \&
  {Crawford}}]{lawrence99}
{Lawrence} A. {et~al.}, 1999, MNRAS, 308, 897

\bibitem[{{Loeb} \& {Narayan}(2008)}]{loeb08}
{Loeb} A., {Narayan} R., 2008, MNRAS, 386, 2221

\bibitem[{{Lu} {et~al}\mbox{.}(1990){Lu}, {Dow}, {Houck}, {Salpeter}, \&
  {Lewis}}]{lu90}
{Lu} N.~Y., {Dow} M.~W., {Houck} J.~R., {Salpeter} E.~E., {Lewis} B.~M., 1990,
  ApJ, 357, 388

\bibitem[{{Lu} \& {Freudling}(1995)}]{lu95}
{Lu} N.~Y., {Freudling} W., 1995, \apj, 449, 527

\bibitem[{{Martin} {et~al}\mbox{.}(2010){Martin}, {Papastergis}, {Giovanelli},
  {Haynes}, {Springob}, \& {Stierwalt}}]{Martin+2010}
{Martin} A.~M., {Papastergis} E., {Giovanelli} R., {Haynes} M.~P., {Springob}
  C.~M., {Stierwalt} S., 2010, ApJ, 723, 1359

\bibitem[{{Martin} {et~al}\mbox{.}(1990){Martin}, {Bottinelli}, {Dennefeld},
  {Fouque}, {Gouguenheim}, \& {Paturel}}]{martin90}
{Martin} J.~M., {Bottinelli} L., {Dennefeld} M., {Fouque} P., {Gouguenheim} L.,
  {Paturel} G., 1990, A\&A, 235, 41

\bibitem[{{Marzke}, {Huchra} \& {Geller}(1996){Marzke}, {Huchra}, \&
  {Geller}}]{marzke96}
{Marzke} R.~O., {Huchra} J.~P., {Geller} M.~J., 1996, \aj, 112, 1803

\bibitem[{{Masters} {et~al}\mbox{.}(2014){Masters}, {Crook}, {Hong}, {Jarrett},
  {Koribalski}, {Macri}, {Springob}, \& {Staveley-Smith}}]{masters14}
{Masters} K.~L., {Crook} A., {Hong} T., {Jarrett} T.~H., {Koribalski} B.~S.,
  {Macri} L., {Springob} C.~M., {Staveley-Smith} L., 2014, \mnras, 443, 1044

\bibitem[{{Masters}, {Springob} \& {Huchra}(2008){Masters}, {Springob}, \&
  {Huchra}}]{Masters+2008}
{Masters} K.~L., {Springob} C.~M., {Huchra} J.~P., 2008, AJ, 135, 1738

\bibitem[{{Matthews} \& {van Driel}(2000)}]{matthews00}
{Matthews} L.~D., {van Driel} W., 2000, A\&AS, 143, 421

\bibitem[{{McGaugh} {et~al}\mbox{.}(2000){McGaugh}, {Schombert}, {Bothun}, \&
  {de Blok}}]{mcgaugh00}
{McGaugh} S.~S., {Schombert} J.~M., {Bothun} G.~D., {de Blok} W.~J.~G., 2000,
  \apjl, 533, L99

\bibitem[{{Meyer} {et~al}\mbox{.}(2004){Meyer}, {Zwaan}, {Webster},
  {Staveley-Smith}, {Ryan-Weber}, {Drinkwater}, {Barnes}, {Howlett},
  {et~al.}}]{Meyer2004}
{Meyer} M.~J. {et~al.}, 2004, MNRAS, 350, 1195

\bibitem[{{Monnier Ragaigne} {et~al}\mbox{.}(2003){Monnier Ragaigne}, {van
  Driel}, {Schneider}, {Balkowski}, \& {Jarrett}}]{monnier03c}
{Monnier Ragaigne} D., {van Driel} W., {Schneider} S.~E., {Balkowski} C.,
  {Jarrett} T.~H., 2003, A\&A, 408, 465

\bibitem[{{Motch} {et~al}\mbox{.}(1998){Motch}, {Guillout}, {Haberl},
  {Krautter}, {Pakull}, {Pietsch}, {Reinsch}, {Voges}, \& {Zickgraf}}]{motch98}
{Motch} C. {et~al.}, 1998, A\&AS, 132, 341

\bibitem[{{Nakanishi} {et~al}\mbox{.}(1997){Nakanishi}, {Takata}, {Yamada},
  {Takeuchi}, {Shiroya}, {Miyazawa}, {Watanabe}, \& {Saito}}]{nakanishi97}
{Nakanishi} K., {Takata} T., {Yamada} T., {Takeuchi} T.~T., {Shiroya} R.,
  {Miyazawa} M., {Watanabe} S., {Saito} M., 1997, \apjs, 112, 245

\bibitem[{{O'Neil}(2004)}]{oneil04HI}
{O'Neil} K., 2004, AJ, 128, 2080

\bibitem[{{Oosterloo}, {Verheijen} \& {van Cappellen}(2010){Oosterloo},
  {Verheijen}, \& {van Cappellen}}]{oosterloo10}
{Oosterloo} T., {Verheijen} M., {van Cappellen} W., 2010, in ISKAF2010 Science
  Meeting, p.~43

\bibitem[{{Pantoja} {et~al}\mbox{.}(1997){Pantoja}, {Altschuler}, {Giovanardi},
  \& {Giovanelli}}]{pantoja97}
{Pantoja} C.~A., {Altschuler} D.~R., {Giovanardi} C., {Giovanelli} R., 1997,
  AJ, 113, 905

\bibitem[{{Parisi} {et~al}\mbox{.}(2009){Parisi}, {Masetti},
  {Jim{\'e}nez-Bail{\'o}n}, {Chavushyan}, \& {Malizia}}]{parisi09}
{Parisi} P., {Masetti} N., {Jim{\'e}nez-Bail{\'o}n} E., {Chavushyan} V.,
  {Malizia} A., 2009, A\&A, 507, 1345

\bibitem[{{Paturel} {et~al}\mbox{.}(2003){Paturel}, {Theureau}, {Bottinelli},
  {Gouguenheim}, {Coudreau-Durand}, {Hallet}, \& {Petit}}]{paturel03}
{Paturel} G., {Theureau} G., {Bottinelli} L., {Gouguenheim} L.,
  {Coudreau-Durand} N., {Hallet} N., {Petit} C., 2003, A\&A, 412, 57

\bibitem[{{Radburn-Smith} {et~al}\mbox{.}(2006){Radburn-Smith}, {Lucey},
  {Woudt}, {Kraan-Korteweg}, \& {Watson}}]{radburn2006}
{Radburn-Smith} D.~J., {Lucey} J.~R., {Woudt} P.~A., {Kraan-Korteweg} R.~C.,
  {Watson} F.~G., 2006, \mnras, 369, 1131

\bibitem[{{Ramatsoku}(2012)}]{ramatsoku12}
{Ramatsoku} M., 2012, {HI-line mapping of large-scale structures in the Zone of
  Avoidance, Masters Thesis,}. University of Cape Town

\bibitem[{{Ramatsoku} {et~al}\mbox{.}(2014){Ramatsoku}, {Kraan-Korteweg},
  {Schr{\"o}der}, \& {van Driel}}]{ramatsoku14}
{Ramatsoku} M., {Kraan-Korteweg} R., {Schr{\"o}der} A., {van Driel} W., 2014,
  ArXiv e-prints

\bibitem[{{Ramatsoku} {et~al}\mbox{.}(2016){Ramatsoku}, {Verheijen},
  {Kraan-Korteweg}, {J{\'o}zsa}, {Schr{\"o}der}, {Jarrett}, {Elson}, {van
  Driel}, {de Blok}, \& {Henning}}]{ramatsoku16}
{Ramatsoku} M. {et~al.}, 2016, \mnras, 460, 923

\bibitem[{Riad(2010)}]{Riad2010}
Riad I.~F., 2010, PhD thesis, University of Cape Town

\bibitem[{{Riad}, {Kraan-Korteweg} \& {Woudt}(2010){Riad}, {Kraan-Korteweg}, \&
  {Woudt}}]{Riad2010b}
{Riad} I.~F., {Kraan-Korteweg} R.~C., {Woudt} P.~A., 2010, MNRAS, 401, 924

\bibitem[{{Ricci} {et~al}\mbox{.}(2011){Ricci}, {Walter}, {Courvoisier}, \&
  {Paltani}}]{ricci11}
{Ricci} C., {Walter} R., {Courvoisier} T.~J.-L., {Paltani} S., 2011, \aap, 532,
  A102

\bibitem[{{Rosenberg} \& {Schneider}(2000)}]{rosenberg00}
{Rosenberg} J.~L., {Schneider} S.~E., 2000, ApJS, 130, 177

\bibitem[{{Ryan-Weber} {et~al}\mbox{.}(2002){Ryan-Weber}, {Koribalski},
  {Staveley-Smith}, {Jerjen}, {Kraan-Korteweg}, \& {Ryder}}]{ryan-weber02}
{Ryan-Weber} E., {Koribalski} B.~S., {Staveley-Smith} L., {Jerjen} H.,
  {Kraan-Korteweg} R.~C., {Ryder} S.~D., 2002, AJ, 124, 1954

\bibitem[{{Said}, {Kraan-Korteweg} \& {Jarrett}(2015){Said}, {Kraan-Korteweg},
  \& {Jarrett}}]{Said15TF}
{Said} K., {Kraan-Korteweg} R.~C., {Jarrett} T.~H., 2015, \mnras, 447, 1618

\bibitem[{{Said} {et~al}\mbox{.}(2016{\natexlab{a}}){Said}, {Kraan-Korteweg},
  {Jarrett}, {Staveley-Smith}, \& {Williams}}]{Said16NIR}
{Said} K., {Kraan-Korteweg} R.~C., {Jarrett} T.~H., {Staveley-Smith} L.,
  {Williams} W.~L., 2016{\natexlab{a}}, \mnras, 462, 3386

\bibitem[{{Said} {et~al}\mbox{.}(2016{\natexlab{b}}){Said}, {Kraan-Korteweg},
  {Staveley-Smith}, {Williams}, {Jarrett}, \& {Springob}}]{Said16HI}
{Said} K., {Kraan-Korteweg} R.~C., {Staveley-Smith} L., {Williams} W.~L.,
  {Jarrett} T.~H., {Springob} C.~M., 2016{\natexlab{b}}, \mnras, 457, 2366

\bibitem[{{Saintonge}(2007)}]{saintonge07}
{Saintonge} A., 2007, AJ, 133, 2087

\bibitem[{{Saunders} {et~al}\mbox{.}(1991){Saunders}, {Frenk},
  {Rowan-Robinson}, {Efstathiou}, {Lawrence}, {Kaiser}, {Ellis}, {Crawford},
  {Xia}, \& {Parry}}]{saunders1991}
{Saunders} W. {et~al.}, 1991, \nat, 349, 32

\bibitem[{{Saunders} {et~al}\mbox{.}(2000){Saunders}, {Sutherland}, {Maddox},
  {Keeble}, {Oliver}, {Rowan-Robinson}, {McMahon}, {Efstathiou}, {Tadros},
  {White}, {Frenk}, {Carrami{\~n}ana}, \& {Hawkins}}]{saunders00}
{Saunders} W. {et~al.}, 2000, MNRAS, 317, 55

\bibitem[{{Saurer}, {Seeberger} \& {Weinberger}(1997){Saurer}, {Seeberger}, \&
  {Weinberger}}]{saurer1997}
{Saurer} W., {Seeberger} R., {Weinberger} R., 1997, \aaps, 126, 247

\bibitem[{{Schlafly} \& {Finkbeiner}(2011)}]{schlafly11}
{Schlafly} E.~F., {Finkbeiner} D.~P., 2011, \apj, 737, 103

\bibitem[{{Schlegel}, {Finkbeiner} \& {Davis}(1998){Schlegel}, {Finkbeiner}, \&
  {Davis}}]{schlegel98}
{Schlegel} D.~J., {Finkbeiner} D.~P., {Davis} M., 1998, ApJ, 500, 525

\bibitem[{{Schneider} {et~al}\mbox{.}(1986){Schneider}, {Helou}, {Salpeter}, \&
  {Terzian}}]{schneider86}
{Schneider} S.~E., {Helou} G., {Salpeter} E.~E., {Terzian} Y., 1986, AJ, 92,
  742

\bibitem[{{Schneider} {et~al}\mbox{.}(1990){Schneider}, {Thuan}, {Magri}, \&
  {Wadiak}}]{schneider90}
{Schneider} S.~E., {Thuan} T.~X., {Magri} C., {Wadiak} J.~E., 1990, ApJS, 72,
  245

\bibitem[{{Schr{\"o}der}, {Kraan-Korteweg} \& {Henning}(2009){Schr{\"o}der},
  {Kraan-Korteweg}, \& {Henning}}]{Schroder2009}
{Schr{\"o}der} A.~C., {Kraan-Korteweg} R.~C., {Henning} P.~A., 2009, A\&A, 505,
  1049

\bibitem[{{Schr{\"o}der}, {van Driel} \& {Kraan-Korteweg}(in
  prep.){Schr{\"o}der}, {van Driel}, \& {Kraan-Korteweg}}]{Schroeder2018}
{Schr{\"o}der} A.~C., {van Driel} W., {Kraan-Korteweg} R.~C., in prep., \mnras

\bibitem[{{Scrimgeour} {et~al}\mbox{.}(2016){Scrimgeour}, {Davis}, {Blake},
  {Staveley-Smith}, {Magoulas}, {Springob}, {Beutler}, {Colless}, {Johnson},
  {Jones}, {Koda}, {Lucey}, {Ma}, {Mould}, \& {Poole}}]{scrimgeour2016}
{Scrimgeour} M.~I. {et~al.}, 2016, \mnras, 455, 386

\bibitem[{{Seeberger}, {Huchtmeier} \& {Weinberger}(1994){Seeberger},
  {Huchtmeier}, \& {Weinberger}}]{seeberger94}
{Seeberger} R., {Huchtmeier} W.~K., {Weinberger} R., 1994, A\&A, 286, 17

\bibitem[{{Seeberger} \& {Saurer}(1998)}]{seeberger98}
{Seeberger} R., {Saurer} W., 1998, A\&As, 127, 101

\bibitem[{{Skrutskie} {et~al}\mbox{.}(2006){Skrutskie}, {Cutri}, {Stiening},
  {Weinberg}, {Schneider}, {Carpenter}, {Beichman}, {Capps}, {Chester},
  {Elias}, {Huchra}, {Liebert}, {Lonsdale}, {Monet}, {Price}, {Seitzer},
  {Jarrett}, {Kirkpatrick}, {Gizis}, {Howard}, {Evans}, {Fowler}, {Fullmer},
  {Hurt}, {Light}, {Kopan}, {Marsh}, {McCallon}, {Tam}, {Van Dyk}, \&
  {Wheelock}}]{skrutskie06}
{Skrutskie} M.~F. {et~al.}, 2006, \aj, 131, 1163

\bibitem[{{Spinrad}(1975)}]{spinrad75}
{Spinrad} H., 1975, \apjl, 199, L1

\bibitem[{{Spinrad} {et~al}\mbox{.}(1985){Spinrad}, {Marr}, {Aguilar}, \&
  {Djorgovski}}]{spinrad85}
{Spinrad} H., {Marr} J., {Aguilar} L., {Djorgovski} S., 1985, PASP, 97, 932

\bibitem[{{Springob} {et~al}\mbox{.}(2005){Springob}, {Haynes}, {Giovanelli},
  \& {Kent}}]{springob05}
{Springob} C.~M., {Haynes} M.~P., {Giovanelli} R., {Kent} B.~R., 2005, ApJS,
  160, 149

\bibitem[{{Springob} {et~al}\mbox{.}(2016){Springob}, {Hong}, {Staveley-Smith},
  {Masters}, {Macri}, {Koribalski}, {Jones}, {Jarrett}, {Magoulas}, \& {Erdo{\u
  g}du}}]{springob2016}
{Springob} C.~M. {et~al.}, 2016, \mnras, 456, 1886

\bibitem[{{Springob} {et~al}\mbox{.}(2014){Springob}, {Magoulas}, {Colless},
  {Mould}, {Erdo{\u g}du}, {Jones}, {Lucey}, {Campbell}, \&
  {Fluke}}]{springob2014}
{Springob} C.~M. {et~al.}, 2014, \mnras, 445, 2677

\bibitem[{{Staveley-Smith} {et~al}\mbox{.}(2016){Staveley-Smith},
  {Kraan-Korteweg}, {Schr{\"o}der}, {Henning}, {Koribalski}, {Stewart}, \&
  {Heald}}]{hizoa2016}
{Staveley-Smith} L., {Kraan-Korteweg} R.~C., {Schr{\"o}der} A.~C., {Henning}
  P.~A., {Koribalski} B.~S., {Stewart} I.~M., {Heald} G., 2016, \aj, 151, 52

\bibitem[{{Strauss} {et~al}\mbox{.}(1992){Strauss}, {Huchra}, {Davis}, {Yahil},
  {Fisher}, \& {Tonry}}]{strauss92}
{Strauss} M.~A., {Huchra} J.~P., {Davis} M., {Yahil} A., {Fisher} K.~B.,
  {Tonry} J., 1992, ApJS, 83, 29

\bibitem[{{Takata} {et~al}\mbox{.}(1994){Takata}, {Yamada}, {Saito},
  {Chamaraux}, \& {Kazes}}]{takata94}
{Takata} T., {Yamada} T., {Saito} M., {Chamaraux} P., {Kazes} I., 1994, A\&AS,
  104, 529

\bibitem[{{Tempel} {et~al}\mbox{.}(2014){Tempel}, {Stoica}, {Mart{\'{\i}}nez},
  {Liivam{\"a}gi}, {Castellan}, \& {Saar}}]{tempel2014}
{Tempel} E., {Stoica} R.~S., {Mart{\'{\i}}nez} V.~J., {Liivam{\"a}gi} L.~J.,
  {Castellan} G., {Saar} E., 2014, \mnras, 438, 3465

\bibitem[{{Tully} {et~al}\mbox{.}(2008){Tully} {et~al.}}]{Tully2008}
{Tully} R.~B., {et~al.}, 2008, ApJ, 676, 184

\bibitem[{{van Driel} {et~al}\mbox{.}(2016){van Driel}, {Butcher}, {Schneider},
  {Lehnert}, {Minchin}, {Blyth}, {Chemin}, {Hallet}, {Joseph}, {Kotze},
  {Kraan-Korteweg}, {Olofsson}, \& {Ramatsoku}}]{vandriel16}
{van Driel} W. {et~al.}, 2016, \aap, 595, A118

\bibitem[{{van Driel} {et~al}\mbox{.}(2009){van Driel}, {Schneider},
  {Kraan-Korteweg}, \& {Monnier Ragaigne}}]{vandriel09}
{van Driel} W., {Schneider} S.~E., {Kraan-Korteweg} R.~C., {Monnier Ragaigne}
  D., 2009, A\&A, 505, 29

\bibitem[{{Visvanathan} \& {Yamada}(1996)}]{visvanathan96}
{Visvanathan} N., {Yamada} T., 1996, ApJS, 107, 521

\bibitem[{{Wakamatsu} {et~al}\mbox{.}(1994){Wakamatsu}, {Hasegawa}, {Karoji},
  {Sekiguchi}, {Menzies}, \& {Malkan}}]{Wakamatsu1994}
{Wakamatsu} K., {Hasegawa} T., {Karoji} H., {Sekiguchi} K., {Menzies} J.~W.,
  {Malkan} M., 1994, in Astronomical Society of the Pacific Conference Series,
  Vol.~67, Unveiling Large-Scale Structures Behind the Milky Way, {Balkowski}
  C., {Kraan-Korteweg} R.~C., eds., p. 131

\bibitem[{{Wakamatsu} {et~al}\mbox{.}(2005){Wakamatsu}, {Malkan}, {Nishida},
  {Parker}, {Saunders}, \& {Watson}}]{Wakamatsu2005}
{Wakamatsu} K., {Malkan} M.~A., {Nishida} M.~T., {Parker} Q.~A., {Saunders} W.,
  {Watson} F.~G., 2005, in Astronomical Society of the Pacific Conference
  Series, Vol. 329, Nearby Large-Scale Structures and the Zone of Avoidance,
  {Fairall} A.~P., {Woudt} P.~A., eds., p. 189

\bibitem[{{Wilson}, {Rohlfs} \& {Huttemeister}(2009){Wilson}, {Rohlfs}, \&
  {Huttemeister}}]{wilson09}
{Wilson} T.~L., {Rohlfs} K., {Huttemeister} S., 2009, {Tools of Radio
  Astronomy}. Springer-Verlag

\bibitem[{{Wong} {et~al}\mbox{.}(2006){Wong}, {Ryan-Weber}, {Garcia-Appadoo},
  {Webster}, {Staveley-Smith}, {Zwaan}, {Meyer}, {Barnes}, {et~al.}}]{Wong2006}
{Wong} O.~I. {et~al.}, 2006, MNRAS, 371, 1855

\bibitem[{{Woudt} {et~al}\mbox{.}(2008){Woudt}, {Kraan-Korteweg}, {Lucey},
  {Fairall}, \& {Moore}}]{woudt2008}
{Woudt} P.~A., {Kraan-Korteweg} R.~C., {Lucey} J., {Fairall} A.~P., {Moore}
  S.~A.~W., 2008, \mnras, 383, 445

\bibitem[{{Yamada} {et~al}\mbox{.}(1993){Yamada}, {Takata}, {Djamaluddin},
  {Tomita}, {Aoki}, {Takeda}, \& {Saito}}]{yamada93}
{Yamada} T., {Takata} T., {Djamaluddin} T., {Tomita} A., {Aoki} K., {Takeda}
  A., {Saito} M., 1993, \apjs, 89, 57

\bibitem[{{Yamada} {et~al}\mbox{.}(1994){Yamada}, {Tomita}, {Saito},
  {Chamaraux}, \& {Kazes}}]{yamada94}
{Yamada} T., {Tomita} A., {Saito} M., {Chamaraux} P., {Kazes} I., 1994, \mnras,
  270, 93

\bibitem[{{Zwaan} {et~al}\mbox{.}(2005){Zwaan}, {Meyer}, {Staveley-Smith}, \&
  {Webster}}]{Zwaan+2005}
{Zwaan} M.~A., {Meyer} M.~J., {Staveley-Smith} L., {Webster} R.~L., 2005,
  MNRAS, 359, L30

\bibitem[{{Zwaan} {et~al}\mbox{.}(2003){Zwaan}, {Staveley-Smith}, {Koribalski},
  {Henning}, {Kilborn}, {Ryder}, {Barnes}, {Bhathal}, {Boyce}, {de Blok},
  {Disney}, {Drinkwater}, {Ekers}, {Freeman}, {Gibson}, {Green}, {Haynes},
  {Jerjen}, {Juraszek}, {Kesteven}, {Knezek}, {Kraan-Korteweg}, {Mader},
  {Marquarding}, {Meyer}, {Minchin}, {Mould}, {O'Brien}, {Oosterloo}, {Price},
  {Putman}, {Ryan-Weber}, {Sadler}, {Schr{\"o}der}, {Stewart}, {Stootman},
  {Warren}, {Waugh}, {Webster}, \& {Wright}}]{zwaan03}
{Zwaan} M.~A. {et~al.}, 2003, AJ, 125, 2842

\end{thebibliography}
